\documentclass[english,aps, superscriptaddress, a4, 12pt, showpacs]{revtex4}
\pdfoutput=1

\usepackage{graphicx}

\usepackage{amssymb}
\usepackage{amsmath}
\usepackage{babel}

\newcommand{\be}{\begin{eqnarray}}
\newcommand{\ee}{\end{eqnarray}}
\newcommand{\nn}{\nonumber }

\usepackage{babel}
\makeatother

\begin{document}

\title{Finite-Size Scaling behavior in the $O(4)$-Model}
\author{Jens~Braun}
\affiliation{TRIUMF, 4004 Wesbrook Mall, Vancouver, BC V6T 2A3, Canada}
\author{Bertram~Klein}
\affiliation{Technische Universit\"at M\"unchen, James-Franck-Strasse 1, 85747 Garching, Germany}
\date{\today}                   

\begin{abstract}
The exact nature of the QCD phase transition has still not been determined conclusively, and there are contradictory 
results from lattice QCD simulations about the scaling behavior for two quark flavors. Ultimately, this issue 
can be resolved only by a careful scaling and finite-size scaling analysis of the lattice results. We use a renormalization group 
approach to obtain finite-size scaling functions for the $O(4)$-model, which are relevant for this analysis. Our results are applicable to Lattice QCD studies of
the QCD phase boundary.
\end{abstract}


\maketitle

\section{Introduction}
Quantum Chromodynamics (QCD) at finite temperature and density is currently actively researched both on the
experimental and the theoretical side. The equation of state of QCD and, in particular, the nature of the
phase transition from the strongly interacting hadronic phase to the strongly interacting quark-gluon plasma
phase is of great importance for a better understanding of the experimental data~\cite{BraunMunzinger:2003zd}.

In QCD, two phase transitions take place at finite temperature and density: a deconfinement phase transition
dominated by the gauge fields and a chiral phase transition driven by the interplay between quark
and gauge-field degrees of freedom. 
The deconfinement transition in pure SU(3) gauge theory is of first order, but becomes a crossover in the presence of dynamical quarks. The nature of the chiral phase transition, in turn, depends on the number of quark flavors, the strength of the explicit chiral symmetry breaking, and the strength of the chiral anomaly~\cite{Pisarski:1983ms,Chandrasekharan:2007up}.  
Whether the chiral and the deconfinement
phase transition coincide is still under investigation. 
Assuming that the effects
of the anomaly at the chiral phase boundary are small, one expects a second order phase transition for two quark flavors in the chiral limit and that QCD falls into the $O(4)$ universality class~\cite{Pisarski:1983ms}. In this case the phase transition in QCD would be dominated by the restoration of chiral symmetry.
With explicit symmetry breaking, the order of the phase transition changes
instantaneously from second order to a crossover. 
If the strength of the chiral anomaly is large, however, for two massless quark flavors one expects that the transition is of first order~\cite{Pisarski:1983ms,Chandrasekharan:2007up}.

For studying full QCD, lattice simulations are currently the most powerful tool. However, the determination of the
nature of the chiral phase transition  still remains a difficult task, since such simulations are 
necessarily performed in finite volumes and the implementation of chiral fermions is difficult.
While there is much evidence that QCD with two flavors falls into the $O(4)$ universality class~\cite{Aoki:2006we}, 
results obtained with two dynamical flavors of staggered fermions do not exhibit the expected scaling behavior \cite{Aoki:1998wg,Bernard:1999fv,Engels:2005rr} or suggest a first-order transition~\cite{D'Elia:2005bv,Cossu:2007mn}.

The finite simulation volume of lattice simulations poses in particular a problem for the investigation of phase transitions. 
Since phase transitions occur strictly speaking only in the infinite-volume limit 
and a continuous symmetry cannot be spontaneously broken in a finite volume, the
introduction of explicit symmetry breaking in the form of a finite current quark mass term 
is mandatory. This makes it difficult to determine the nature of the chiral phase transition from 
lattice QCD results. An important tool for the analysis of lattice QCD data is the investigation of the  finite-size 
scaling behavior. The underlying universality class determines the 
scaling behavior of e.g. the order parameter characterizing the transition and the corresponding susceptibility.
If universal behavior obtains, results are expected to fall onto universal scaling 
curves, characterized by critical exponents and scaling functions of the underlying universality class. 
Thus the order of the phase transition and its universality class can be established by a comparison with the known critical exponents and the known scaling behavior. From such an analysis, there is indeed evidence of 
$O(4)$ scaling for QCD with Wilson fermions~\cite{AliKhan:2000iz,Mendes:2006zf,Mendes:2007ve,Ejiri:2007qk}, however, the expected scaling behavior has not been
seen with staggered fermions \cite{Aoki:1998wg,Bernard:1999fv,Engels:2005rr}. Results with a modified QCD action with two flavors of staggered
fermions suggest that current simulation volumes might be actually outside the finite-size scaling region~\cite{Kogut:2006gt}. 
This result underlines the importance of a finite-volume scaling analysis of $O(N)$ models.

So far, scaling and finite-size scaling functions have been determined mainly from lattice 
simulations of $O(N)$ spin models~\cite{Engels:2001bq,Schulze:2001zg,Ballesteros:1996bd}.
These results have already been used in the scaling analysis of Lattice QCD results \cite{Mendes:2006zf,Mendes:2007ve,Engels:2001bq,Engels:2005rr,Engels:2005te}. 
In this paper, we provide the technical framework for finite-size scaling studies with the 
functional 
renormalization group (RG) and 
compute the finite-size scaling functions for the 
linear
$O(4)$ model.  
Our RG approach complements the one taken in $O(N)$ lattice simulations.
It is computationally efficient and 
allows us to study scaling over a wide range of volume sizes and values of the external symmetry breaking field, which is usually associated with an external
magnetic field in spin models or, in the case of QCD, with the current quark mass.
The results are directly applicable to a comparison with Lattice QCD data~\cite{BraunKlein:2008}. 
Making use of universality arguments, our approach enables us 
to study the scaling of the order parameter and the associated susceptibilities at the phase boundary where the 
fermions are assumed to have decoupled from the critical fluctuations. While the approach in the present paper cannot be used to study the onset of chiral
symmetry breaking in terms of quark-gluon dynamics~\cite{Braun:2005uj,Braun:2006jd,Braun:2008pi}, it still provides important information for an analysis of Lattice QCD data which helps to shed light on the ongoing discussion about the nature of the QCD phase transition.

The paper is organized as follows: In Sect.~\ref{sec:FV_RG}, we present the setup of our RG formalism.
In Sects.~\ref{sec:FS_scaling} and~\ref{sec:det_fs_scaling_fcts}, we discuss general aspects of finite-volume and finite-size scaling in
quantum field theories. Our results are then presented in Sec.~\ref{sec:results}. Concluding remarks and future plans
are given in Sect.~\ref{sec:conclusions}.

\section{Renormalization Group Approach for finite volume studies}
\label{sec:FV_RG}
In this section we discuss our Renormalization Group 
approach to studying
finite-volume scaling. In the first part of this section, we briefly discuss the derivation of the
flow equations for the $O(4)$-potential in infinite volume. A detailed discussion of the 
derivation and the underlying approximations can be found in Ref.~\cite{Braun:2007td}.
In the second part of this section we generalize our flow equations
 to finite-volume and recapitulate some earlier results \cite{Braun:2004yk,Braun:2005gy,Braun:2005fj}.
\subsection{Infinite volume}
The effective action of the $O(4)$ model in $d=3$ spatial 
dimensions is given by
\begin{equation}
\Gamma [\phi]=
\int d^{3}x \left\{
\frac{1}{2}(\partial_{\mu}\phi)^{2}+ U(\sigma,\vec{\pi}^2)\right\},
\end{equation}
where $\phi^T=(\sigma, \vec{\pi})$. The potential $U(\sigma,\vec{\pi} ^2)$ depends on $\sigma$ and $\vec{\pi}^2$ separately, since
the presence of a non-vanishing external source term ($-H\sigma$) in the
ansatz for the effective action is indispensable for a study of phase transitions in finite volume.
We study the $O(4)$ model in the so-called local potential approximation (LPA), where we neglect a possible space dependence of the expectation value $\langle\phi\rangle$
and take the wave-function renormalization $Z_{\phi}$
to be constant, $Z_{\phi}=1$. Since the anomalous dimension associated with $Z_{\phi}$ 
is small compared to one, see e. g. Ref.~\cite{Tetradis:1993ts}, 
our approximation, in which the running of the wave-function renormalization is neglected, is
well justified for a first study of finite-size scaling.
The components of the vector $\phi$ are labeled according the role the
corresponding fields are playing in the spontaneously broken regime, $\phi^{\mathrm{{T}}}
=(\sigma,\pi^{1},\pi^{2},\pi^{2})$. We choose the first component to be the radial
mode in the regime where the ground state of the theory is not symmetric 
under $O(4)$ transformations:
\be
\langle \phi \rangle = \phi_{0} ^{\mathrm{{T}}} =(\sigma_0,0,0,0)\,.
\ee

The RG flow equation for the effective action according to C.~Wetterich~\cite{Wetterich:1992yh} reads:
\be
\partial _t \Gamma_k =\frac{1}{2}\text{STr}\,(\partial_t R_k)\cdot
\left[\Gamma_k ^{(2)}+R_k\right]^{-1}\,,\label{eq:FlowEq}
\ee
where the dimensionless flow variable $t$ is given by $t=\ln (k/\Lambda)$ and $\Lambda$ denotes
a UV cutoff at which all couplings are initially specified. The regulator function $R_k$ specifies the details of the Wilsonian momentum-shell
integrations and has to satisfy certain constraints~\cite{Wetterich:1992yh}. Since the choice of the 
regulator function is at our disposal, we can use it to optimize the RG flow~\cite{Litim:2000ci,Litim:2001fd,Litim:2001up,Pawlowski:2005xe}.
In the following, we employ the optimized regulator function~\cite{Litim:2001up}
\be
R_k(p^2)=p^2 r(p^2/k^2)\qquad\text{with}\qquad r(x)=\left(\frac{1}{x}-1\right)\Theta(1-x)\,.\label{eq:regulator}
\ee
We then find for the flow equation for the effective potentials~\cite{Litim:2001up,Litim:2001hk}:
\begin{eqnarray}
k \frac{\partial}{\partial k} U_k &=&   \frac{k^5}{(4 \pi)^{\frac{3}{2}}}\frac{1}{\Gamma(\frac{3}{2}+1)} \left( \frac{3}{k^2 + M^2_{\pi,k}} 
+ \frac{1}{k^2 +M^2_{\sigma,k}} \right).
\label{eq:floweq_dim}
\end{eqnarray}
Here we have replaced the bare masses and couplings in the
inverse two-point functions with the scale-dependent quantities.
The quantities $M_{\sigma}$ and $M_{\pi}$ are the eigenvalues of the 
the second-derivative matrix of the potential. Note that
these quantities still depend on the background fields $\sigma$ and $\vec{\pi}^2$.

In Refs.~\cite{Braun:2004yk,Braun:2005gy,Braun:2005fj}, the proper-time Renormalizaton
Group (PTRG) has been used to study finite-volume effects in a quark-meson model.
Within the PTRG framework, the 
RG flow equations for the potential can be derived straighfowardly by 
inserting a cutoff function into the Laplace-transform of the one-loop
effective action ~\cite{Liao:1994cm, Schaefer:1999em,Litim:2001up,Litim:2001hk, Bohr:2000gp,Braun:2007td}.
Although the PTRG framework does not, in general, yield exact RG flow equations~\cite{Litim:2001hk,Litim:2001ky}, the PTRG cutoff function can be
chosen such that the resulting flow equations for the effective potential in LPA are identical
to Eq.~\eqref{eq:floweq_dim}, which was found in Refs.~\cite{Litim:2001up,Litim:2001hk}.

For studying scaling behavior it is convenient
to deal with dimensionless quantities rather than dimensionful
quantities. Therefore we introduce the dimensionless potential $u$, the
dimensionless masses $m_{\sigma}$ and $m_{\pi}$, as well as 
the dimensionless field-vector $\varphi$ by
\be
u_k =k^{-3} U_k \,\quad m^2 _{i,k} =k^{-2} M^2 _{i,k}
\quad\text{and}\quad \varphi_i=k^{-\frac{1}{2}}\phi_i\,.
\label{eq:dimquant}
\ee
Applying these definitions to the flow equation~\eqref{eq:floweq_dim}, we obtain
\be
\partial_t u &=&  -3 u  + \frac{1}{(4 \pi)^{3/2}}\frac{1}{\Gamma(3/2+1)} \left( \frac{3}{1 
+ m^2 _{\pi}} + \frac{1}{1 + m^2_{\sigma}} \right)\,,
\label{eq:floweq_dimless}
\ee
Integrating the flow equation
from the UV scale $\Lambda$ to $k \to 0$, we obtain an effective potential in which quantum 
corrections from all scales have been systematically included.

Since we are eventually interested in phase transitions in finite volume
we need a linear term with a source term $H$ in the ansatz for the effective action, which corresponds to an external magnetic field in a spin model. In order to solve the RG flow for the effective potential $U$ (or $u$), we
expand the potential in local $n$-point couplings around its minimum $\sigma_0(k)$
\be
U_k(\sigma, \vec{\pi}^2) = a_0(k) + a_1(k) (\sigma^2 + \vec{\pi}^2 - \sigma_0(k)^2) + a_2(k) (\sigma^2 + \vec{\pi}^2 - \sigma_0(k)^2)^2 + \ldots - H \sigma\,,
\label{eq:pot_ansatz}
\ee
where $H$ is the fixed, external symmetry-breaking field and all other couplings and the minimum are scale-dependent. Since we have absorbed the
symmetry-breaking linear term into the ansatz for the potential, $U_k$
depends on the fields $\sigma$ and $\vec{\pi}$ separately. The condition
\be
\frac{\partial U_k(\sigma,\vec{\pi}^2)}{\partial \sigma}\Bigg|_{\sigma=\sigma_0(k),\vec{\pi}^2=0} 
\stackrel{!}{=} 0
\label{eq:min_cond}
\ee
ensures that we are expanding around the actual physical minimum.
From Eq. \eqref{eq:min_cond}, we find that the RG flow of the coupling $a_1(k)$ and the minimum $\sigma_0(k)$ 
are related by the condition
\be
2 a_1(k) \sigma_0(k) = H\,.\label{eq:min_cond2}
\ee 
This condition keeps the minimum at $(\sigma, \vec{\pi}) = (\sigma_0(k), \vec{0})$. The flow
equation of the minimum $\sigma_0(k)$ is thus related to the flow of the coupling $a_1(k)$ in a simple way.

The RG flow equations for the couplings $a_i(k)$ can now be obtained 
straightforwardly by expanding the flow equation \eqref{eq:floweq_dim} around
the minimum $\sigma_0 (k)$ and then projecting it onto the derivative of the ansatz~\eqref{eq:pot_ansatz} with respect to $k$. This procedure results in an infinite set of flow equations for the couplings $a_{i}(k)$. In order to solve the set of equations
for the couplings, we have to truncate our ansatz~\eqref{eq:pot_ansatz} 
for the potential. In the following, we include fluctuations around the minimum up
to eighth order in the fields, i. e. we keep track of the running of the couplings 
$a_1$, $a_2$, $a_3$ and $a_4$. The resulting finite set of coupled
first-order differential equations is then solved numerically. From investigations of the convergence behavior \cite{Tetradis:1993ts,Papp:1999he}, we expect that such a truncation is sufficient for our purpose. Below we confirm explicitly that the scaling functions for the couplings of interest ($\sigma_0$ and $a_2$) satisfy the expected scaling relations, which is strong evidence that the expansion has converged sufficiently and the truncation at this order is justified.
\subsection{Finite volume}
Now we generalize the RG flow equations in the first part of this section to a finite $d=3$ dimensional Euclidean volume. This is done 
by replacing the integrals over the momenta in the evaluation of the trace in Eq.~\eqref{eq:FlowEq} by a sum
\be
\int \frac{dp_{i}}{2\pi}\,\ldots \rightarrow \frac{1}{L} \sum_{n_{i} =
  -\infty}^{\infty}\ldots\,. 
\ee
We only consider isotropic volumes, but the approach is not limited
to these. Anisotropic Euclidean volumes have been used in a 
study of the quark-meson model in $d=4$ dimensions, see Ref.~\cite{Braun:2005fj}.
The boundary conditions in the Euclidean time direction are fixed
by the statistics of the fields, i. e. we must choose periodic boundary
conditions in this direction. In the present case, there are only three spatial dimensions. In order to
be able to compare our results with Lattice simulations of $O(N)$-models, 
we choose periodic boundary conditions in these spatial dimensions. Thus the $3$-momenta are discretized as follows:
\begin{equation}
p^{2}=\frac{4\pi^{2}}{L^{2}} (n_{1}^{2} + n_2^{2} + n_3 ^{3})\,,
\label{eq:fv_mom}
\end{equation}
where $n_i \in \mathbb{Z}_0$ for $i=1,2,3$. The flow equation for the effective potential $u$ can now be derived straightforwardly. From 
the flow equation~\eqref{eq:FlowEq} together with the regulator function~\eqref{eq:regulator}, we
obtain
\be
\partial _t u_t \Big|_{\text{ERG}} &=& -3 u + 3 \tilde{L}_{0}(m_{\pi}^2,kL) + \tilde{L}_{0}(m_{\sigma}^2,kL)\,,\label{eq:FRG_flow_FV}
\ee
where we introduced the finite-volume threshold functions
\be
\tilde{L}_{j}(\omega,x)=\frac{1}{x^{3}}\frac{(j+\delta_{j0})}{(1+\omega)^{j+1}}\sum_{\vec{n}}
\Theta (1-({\textstyle\frac{2\pi}{x}})^{2}\vec{n}^{2})\,.
\ee
The sum counts the number of lattice nodes located in a three-dimensional 
ball with radius $\frac{x}{2\pi}$. In the limit $x\to\infty$ (i. e. $L\to\infty$), the sum can be written as an integral, which yields the well-known threshold functions in three dimensions~\cite{Litim:2001up}:
\be
\tilde{L}_{j}(\omega)=\frac{1}{(4\pi)^{\frac{3}{2}}\Gamma(\frac{3}{2}+1)}\frac{(j+\delta_{j0})}{(1+\omega)^{j+1}}\,.
\ee

In the preceding part of this section, we have pointed out that a PTRG cutoff function can be chosen such
that the resulting PTRG flow equations are identical to those obtained from the flow
equation~\eqref{eq:FlowEq} when the optimized regulator~\eqref{eq:regulator} is employed.
Although this correspondence is true for infinite volume, this is not the case for finite volume. In the
following, we use the same PTRG cutoff function as in the derivation of the flow equation~\eqref{eq:floweq_dimless}
in infinite volume. The flow equation can be derived along the same lines as in Refs.~\cite{Braun:2005fj,Braun:2006zz} and
we find
\be
\partial _t u_t \Big|_{\text{PTRG}} & = & -3 u +
 3\,\Theta_{p}^{(B)} (m_{\pi }^{2},kL) +  \Theta_{p}^{(B)} (m_{\sigma}^{2},kL).\,\label{eq:FV_fe}
\ee
For convenience, we have introduced the (dimensionless) threshold-function
\be
\Theta_{p}^{(B)}(\omega,x)&=&
 \frac{x^{2}}{(4\pi)^{\frac{3}{2}+1}\Gamma(\frac{3}{2}+1)} \int_{0}^{\infty}ds\,
s^{\frac{3}{2}}\mathrm{e}^{-\frac{s(1+\omega) x^2}{4\pi}}
\Big(\vartheta_{p}(s)\Big)^{3}\,,\label{eq:ThetaB} 
\ee
where $\vartheta_{p}$ is the Jacobi-Elliptic-Theta
function defined by
\begin{eqnarray}
\vartheta_{p}(x) &=& \sum_{n=-\infty}^{\infty} \mathrm{e}^{-x\pi n^{2}} =
    x^{-\frac{1}{2}} + 2\sum_{q=1}^{\infty} x^{-\frac{1}{2}}
    \mathrm{e}^{-\frac{\pi q^{2}}{x}}\,.\label{eq:theta_p}  
\end{eqnarray}
The first representation in Eq.~\eqref{eq:theta_p} corresponds to the standard Matsubara summation of the momenta. 
The second representation on the right hand side is obtained by 
applying Poisson's formula to the first representation. Note that the PTRG flow equation~\eqref{eq:FV_fe}
is identical to the flow equation~\eqref{eq:FRG_flow_FV} in the small-volume limit ($L\to 0$), where only
the zero modes contribute, and by construction also in the infinite-volume limit ($L\to\infty$).
In order to obtain the infinite-volume flow equation analytically from the 
corresponding finite-volume flow equation, we use the second representation of the Jacobi-Elliptic-Theta function in Eq.~\eqref{eq:theta_p}. 
This representation actually separates the finite-volume and
the infinite-volume contributions. When we approximate the Jacobi-Elliptic-Theta
function by the first term on the right-hand side of Eq.~\eqref{eq:theta_p} and then
perform the integration over $s$ in Eq.~\eqref{eq:ThetaB}, we do indeed recover
the flow equation~\eqref{eq:floweq_dimless} for infinite volume.

On the one hand, physical quantities should not depend on the choice of the
regularization scheme in the limit $k \rightarrow 0$. One the other hand, investigating the regulator 
dependence of the results in this limit allows us to check their quality in a particular truncation. In this paper, we exploit the difference between the results obtained from
Eq.~\eqref{eq:FRG_flow_FV} and Eq.~\eqref{eq:FV_fe} in order to obtain a theoretical error estimate for our results. In addition, in both schemes the restriction to the local potential approximation introduces an additional systematic error.
\section{Finite-size Scaling}
\label{sec:FS_scaling}
Critical behavior in the vicinity of a critical point is governed by the presence of long-range correlations. A finite volume $V=L^d$ affects the critical scaling behavior if the volume size $L$ becomes comparable to the correlation length $\xi$. According to Fisher's finite-size scaling hypothesis \cite{Fisher:1971ks}, observables in the finite-volume system and the infinite-volume system are then related by a function that depends \emph{only} on the ratio of the infinite-volume correlation length $\xi(t, h, L \to \infty)$ and the linear volume size $L$. 

We use the customary notation and denote the reduced temperature by $t=\frac{(T-T_c)}{T_0}$ with the critical temperature $T_c$, and the external symmetry-breaking field by $h=\frac{H}{H_0}$. The values for $T_c$ and the non-universal normalization constants $T_0$ and $H_0$ which we determined for our parameter choices are given in Tab.~\ref{tab:normalization}. For completeness, we include the values of the critical exponents used in the evaluation in Tab.~\ref{tab:critex}. For a discussion, we refer the reader to Ref.~\cite{Braun:2007td}.

\subsection{Finite-size scaling functions}

Applying the scaling hypothesis for example to the order parameter $M$, the ratio of its values in finite and infinite volume is given by~\cite{Fisher:1971ks} 
\be
\frac{M(t, h, L)}{M(t, h, L \to \infty)} &=& {\mathcal F}\left(\frac{\xi(t, h, L \to \infty)}{L} \right).
\label{eq:scalinghypothesis}
\ee
An RG analysis tells us how the couplings $t$ and $h$ have to be changed in order to keep the system invariant under a change of the length scale. If the volume is finite in all dimensions,  the same critical exponents as for infinite volume govern the scaling behavior.
Since the correlation length behaves as $\xi \sim t^{-\nu}$ in the absence of the field $h$, keeping the ratio of correlation length $\xi$ to system size $L$ constant requires to hold $t L^{1/\nu} = \mathrm{const.}$ Likewise at the critical temperature $t=0$,  we need to keep the combination $h L^{\beta \delta/\nu} = \mathrm{const.}$ in order to preserve the ratio of correlation length and system size. 

This completely specifies the behavior of thermodynamic observables in the vicinity of the critical point in a finite volume system, apart from possible scaling corrections.
For example, the order parameter $M(t, h, L)$ as a function of temperature $t$, external symmetry-breaking field $h$ and volume size is expected to behave as~\cite{Fisher:1971ks}
\be
M(t, h, L) &=& L^{-\beta/\nu} \left[ \tilde{Q}_M( tL^{1/\nu}, hL^{\beta\delta/\nu}) + \frac{1}{L^\omega} \tilde{Q}_M^{(1)}( tL^{1/\nu}, hL^{\beta\delta/\nu}) + \ldots  \right],
\ee
where $\omega$ is the critical exponent associated with the first irrelevant operator in the vicinity of the critical point. The additional terms spoil the finite-size scaling behavior for small volume size. They need to be accounted for explicitly  and removed to isolate the universal finite-size scaling function.

In general, the finite-size scaled order parameter $L^{\beta/\nu} M(t, h, L)$ is a function of the finite-size scaled variables $tL^{1/\nu}$ and $hL^{\beta \delta /\nu}$. In order to analyze the results efficiently in terms of only a single variable, we
parametrize the scaling functions in term of the scaling variable $z=t/h^{1/(\beta \delta)}$. This accounts for the usual critical scaling behavior in infinite volume.  We keep the value of $z$ fixed and use the finite-size scaling variable 
\be
\bar{h} = h L^{(\beta \delta)/\nu}
\label{eq:fssvariabledef}
\ee 
to parametrize the finite-size scaling behavior. The scaling variable $\bar{h}$ is not dimensionless but retains an explicit scale dependence through the length scale $L$. 
While we have not done so in the present analysis, it is possible to remove this explicit non-universal scale as well: Using the correlation length $\xi$ as the natural length scale, we can 
form the dimensionless, universal combination 
\be
\frac{L}{\xi(t, h, L)} = M_\sigma(t, h, L) L, 
\ee
which still depends explicitly on $t$, $h$ (and $L$). This relation can be used to identify a unique length scale by considering the limit $L \to \infty$ and $h \to 0$. In this limit, the correlation length behaves according to   
\be
\xi(t) &=& \frac{1}{C_0} t^{-\nu} \nn\\
M_\sigma(t) &=& C_0 \,t^{\nu}.
\ee
The non-universal normalization constant  $C_0$ has the dimension of an inverse length scale and plays the same role for the "coupling" $L$ as the constants $T_0$ and $H_0$ do for the 
couplings $T$ and $H$. Using this constant, the universal dimensionless combination $M_\sigma(t, L \to \infty)L$ becomes
\be
\frac{L}{\xi(t, L \to \infty)}&=& \frac{C_0 L}{t^{-\nu}} = \ell \, t^\nu, 
\ee
where $\ell = C_0 L$ is the dimensionless length required for a truly universal description of the finite-size scaling behavior.
With our choice of parameters for a UV cutoff scale $\Lambda = 1$ GeV, we find $C_0 = 2.393(2)$ MeV, corresponding to a length scale 
\be
\frac{1}{C_0} &=& 82.510(65) \;\; \mbox{fm}.
\ee
We stress that all explicitely given length scales must be normalized with this factor for a completely scale-independent comparison.

In the following part of this section we review some properties of the finite-size scaling functions and make contact with the scaling results in infinite volume

Provided the finite-volume corrections can be neglected or removed, the finite-size scaled order parameter is given by the universal finite-size scaling function 
\be
L^{\beta/\nu} M(t, h, L) = Q_M(z, h L^{(\beta\delta)/\nu}),
\ee
where the scaling variable $z$ is used to parameterize the dependence on the temperature $t$.
For large volumes, the order parameter must asymptotically approach the infinite-volume scaling limit
\be
\lim_{L \to \infty} M(t, h, L ) = h^{1/\delta} f(z).
\ee
This dictates the behavior of the finite-size scaling function $Q_M(z, h L^{(\beta \delta)/\nu})$ for large values of its second argument, $h L^{(\beta\delta)/\nu}$. In order to recover 
the infinite-volume scaling law, the magnetization must behave as
\be
\lim_{L\to \infty}M(t, h, L)&=& \lim_{L \to \infty} L^{-\beta/\nu} Q_M(z, hL^{(\beta\delta)/\nu}) = \lim_{L \to \infty} L^{-\beta/\nu} \left(h L^{(\beta\delta)/\nu} \right)^{1/\delta} f(z) \nn\\
&=& h^{1/\delta} f(z).
\ee
Therefore we expect to find 
\be
Q_M(z, \bar{h}) \simeq f(z) \bar{h}^{1/\delta}  \quad \mbox{for}\quad \bar{h} \to \infty
\ee
for the scaling function. While this is a very useful check for the consistency of the finite-size scaling results with the infinite-volume scaling behavior, it is only of limited practical 
interest for a finite-size scaling analysis: In the region where the finite-size scaling function displays this behavior, the actual results for the order parameter have already converged to the 
infinite-volume result, and there are no large observable finite-size effects. In this region the results for the order parameter from different volume sizes coincide, apart from possible exponentially
small corrections, but the finite-size scaled results $L^{\beta/\nu} M(t, h, L)$ no longer do, see e.g. Fig.~\ref{fig:z0orderFVscaled} for $z=0$ and Fig.~\ref{fig:zpeakorderFVscaled} for
$z=z_p$. Recall that $z=0$ means that we are sitting at the critical temperature whereas $z=z_p$ means that we are sitting at the peak value of the susceptibility. 

Since the longitudinal susceptibility $\chi$ is given by the derivative of the order parameter with respect to the external symmetry-breaking field $H$, the infinite-volume scaling functions $f(z)$ for the order parameter and $f_\chi(z)$ for the susceptibility are related. In infinite volume, one finds the relationship
\be
\chi(t, h) &=& \frac{\partial M}{\partial H} = \frac{1}{H_0} h^{1/\delta -1} \frac{1}{\delta} \left[ f(z) - \frac{1}{\beta} \frac{z}{h} f^\prime(z) \right] \equiv \frac{1}{H_0} h^{1/\delta -1} f_{\chi}(z).
\ee   
As in infinite volume, the finite-size scaling functions for the order parameter and for the susceptibility are related to each other: Taking the derivative of the finite-size scaled order parameter leads directly to the corresponding expression for the susceptibility.
The finite-size scaling function $Q_M(z, hL^{\beta \delta/\nu})$ depends on two variables $z$ and $\bar{h}$. Taking the derivative with respect to $h$, derivatives with respect to both variables appear, which we denote by
\be
Q_M^{(1, 0)}(z, \bar{h})= \frac{\partial}{\partial z} Q_M(z, \bar{h}) \quad \mathrm{and} \quad  Q_M^{(0, 1)}(z, \bar{h})= \frac{\partial}{\partial \bar{h}} Q_M(z, \bar{h}).
\ee
It is further useful to remember that
\be \quad \frac{d \bar{h}}{dh} = L^{(\beta \delta)/\nu}, \quad \frac{dz}{dh} = - \frac{1}{\beta \delta} \frac{z}{h} .\nn
\ee

We obtain the finite-size scaling function for the susceptibility by taking the derivative of the finite-size scaling function  for the order parameter: 
\be
\frac{\partial }{\partial H} L^{\beta/\nu} M(t, h, L) &=& \frac{1}{H_0} \frac{\partial}{\partial h} Q_M(z, h L^{\beta \delta /\nu} ) = \nonumber\\
&=& \frac{1}{H_0} L^{\beta \delta /\nu} \left[Q_{M}^{(0, 1)}(z, hL^{\beta \delta/\nu}) -\frac{z}{\beta \delta}  \frac{1}{hL^{\beta \delta/\nu}}Q_{M}^{(1, 0)}(z, hL^{\beta \delta/\nu})\right]
\ee
Using the scaling relation $\gamma = \beta(\delta -1)$, one finds that the finite-size scaled susceptibility is given by the scaling function
\be
H_0 L^{-\gamma/\nu} \chi(t, h, L) &=& Q_{M}^{(0, 1)}(z, h L^{\beta \delta /\nu}) - \frac{1}{hL^{\beta \delta /\nu}} \frac{z}{\beta \delta} Q_M^{(1, 0)}(z, h L^{\beta \delta /\nu}) 
\equiv Q_\chi(z, h L^{\beta \delta /\nu}).
\label{eq:suscrelation}
\ee
Since we determine the scaling functions for fixed values of $z$ as a function of the finite-size scaling variable $hL^{\beta \delta /\nu}$, we can test this relationship between the finite-size scaling functions most easily for the case $z=0$, where the second term vanishes. Here we assume 
that the derivative of $Q_M$ with regard to $z$ is bounded for $z \to 0$, or that 
$\lim_{z \to 0} z\; Q^{(1, 0)}_M(z, \bar{h} )= 0$, which is indeed the case in our results.

For large volumes, once again the finite-size scaling function must coincide with the scaling functions in infinite volume, 
\be
\lim_{L \to \infty}\chi(t, h, L)= \lim_{L \to \infty} L^{\gamma/\nu}  \left(hL^{\beta\delta/\nu} \right)^{1/\delta -1} \frac{1}{H_0} f_{\chi}(z) = h^{1/\delta -1} \frac{1}{H_0} f_{\chi}(z), 
\ee
where we used the scaling relation $\beta(\delta -1 ) = \gamma$ again.
Consequently the scaling function must behave asymptotically for large values of $\bar{h}$ as
\be
Q_{\chi}(z, \bar{h}) &=& \bar{h}^{1/\delta -1} \frac{1}{H_0} f_{\chi}(z),
\ee
where the scaling function $f_{\chi}(z)$ and the normalization constant $H_0$ are known from the infinite-volume result.
\subsection{Scaling behavior as a function of the correlation length}
The principal problem in comparing universal scaling functions to actual data is the determination of the non-universal normalization constants. The determination of these scales is essential for a comparison of the scale-free universal results, e. g. the constants in a Lattice QCD simulations are in general different from those in a lattice simulation of an $O(4)$ spin model.
But the determination from a limited data set is difficult, since such a set is often obtained in the presence of strong symmetry-breaking fields 
and in small volumes. These may lead to non-universal scaling corrections where details of the short-range physics enter.

The most useful results are thus those that rely only on the long-range properties of the system in question. Dimensionless variables into which only the actual observables enter and for which any normalization with regard to short-range physics is unnecessary are actually the best candidates for a practical evaluation.

Considering the discussion of finite-size scaling above, the most natural variable for plotting finite-volume results is the ratio of the infinte-volume correlation length to the system size
\[
\frac{\xi(t, h, \infty)}{L}.
\]
But the \emph{infinite-volume} correlation length is itself not directly measurable either, so it is necessary to use the \emph{finite-volume} correlation length, which can be measured in the finite-volume system.

Since the correlation length is an observable, we can apply the finite-size scaling analysis to the correlation length itself. According to the general scaling analysis, for any observable ${\cal O}$, we have up to scaling corrections
\be
{\cal O}(t, h, L) = L^{\kappa/\nu} f_{\cal O}\left(\frac{\xi(t, h, \infty)}{L}\right),
\label{eq:scalingobservable}
\ee
where $\kappa$ is the critical exponent associated with the operator ${\cal O}$. In particular, we have $\kappa=\nu$ for the correlation length itself and the relation then becomes
\be
\frac{\xi(t, h, L)}{L} = f_\xi\left(\frac{\xi(t, h, \infty)}{L}\right).
\ee
Assuming the scaling function in this relationship can be inverted, 
we can express the ratio of the \emph{infinite-volume} correlation length and the volume size as a function of the \emph{finite-volume} correlation length and the volume size:
\be
\frac{\xi(t, h, \infty)}{L} = f_\xi^{-1}\left(\frac{\xi(t, h, L)}{L} \right).
\label{eq:invertedxirelation}
\ee
Substituting this relation into the scaling hypothesis Eq.~\eqref{eq:scalinghypothesis}, the scaling functions can be expressed as a function of the argument $\frac{\xi(t, h, L)}{L} $ involving the finite-volume correlation length. This form is very useful for actual comparisons of different physical systems.

In Sect.~\ref{subsec:FSSF_order}, we will show results for the finite-size scaling function $Q_M(\xi/L)$ for the order parameter at selected values of the scaling 
variable $z$ ($z=0$ and $z=z_p=1.3155$) 
in this form. Since the correlation length and the susceptibility are trivially related in the present approximation, $\chi = 1/M_\sigma^2=\xi^2$, no additional information is gained by determining the finite-size scaling function of the susceptibility as a function of the ratio $\xi/L$.

A different comparison scheme which also eliminates the need for additional normalization constants is based on the comparison of the system for two different volume sizes, $L$ 
and $sL$ with $s\in\mathbb{R}$. This approach is widely used in condensed matter physics, see e.g. Refs.~\cite{Caracciolo:1994ed,Cucchieri:1995yd,Caracciolo:2003nq}. The most 
common choice for the volume ratio is $s=2$.  

Starting from the inverted scaling relation for the finite-volume correlation length Eq.~\eqref{eq:invertedxirelation}, one observes that for a volume of size $sL$
\be
\frac{\xi(t, h, \infty)}{s L} =\frac{1}{s} \frac{\xi(t, h, \infty)}{L} = \frac{1}{s}f_\xi^{-1}\left(\frac{\xi(t, h, L)}{L} \right).
\ee 
Inserting this expression into the scaling relation Eq.~\eqref{eq:scalingobservable} and forming the ratio for system sizes $L$ and $sL$, one obtains
\be
\frac{{\mathcal O}(t, h, sL)}{{\mathcal O}(t, h, L)} &=& s^{\kappa/\nu} \frac{f_{\mathcal O}\left(\frac{1}{s} f_\xi^{-1}\left(\frac{\xi(t, h, L)}{L} \right)\right)}{f_{\mathcal O}\left(f_\xi^{-1}\left(\frac{\xi(t, h, L)}{L} \right)\right)}=:F_{\mathcal O}\left(s, \frac{\xi(t, h, L)}{L} \right).
\ee
Scaling corrections to this relation are of the order $\xi^{-\omega}$ and $L^{-\omega}$, where $\omega$ is the exponent associated with the first irrelevant operator.
The ratio of the values of an observable for two systems sizes $L$ and $sL$ therefore defines a new scaling function $F_{\mathcal O}$ which depends on the ratio $s$ and the dimensionless correlation length $\xi(t, h, L)/L$. This scaling function can be used for direct comparisons without any need for an additional determination of normalization constants. 

In Sect.~\ref{subsec:SR_corrlength}, we discuss the results for this finite-size scaling function for the order parameter and for the correlation length, which corresponds to the 
susceptibility, for selected values of $z$~($z=0$ and $z=z_p=1.3155$).
\section{Determination of the finite-size scaling functions}
\label{sec:det_fs_scaling_fcts}
Ideally, the results for different values of the volume size in the finite-size scaling region would coincide perfectly after rescaling. But as we have already observed for infinite-volume scaling \cite{Braun:2007td}, the external symmetry-breaking field $h$ has to remain small in order to keep corrections to scaling small. This means in particular that the masses of the critical fluctuations must remain very small compared to the UV cutoff scale $\Lambda$. On the other hand, in order to fully explore the finite-size scaling region, we must decrease the correlation length sufficiently to restore the infinite-volume results, which implies an increase in the masses of the fluctuations.  

Therefore corrections to the finite-size scaling behavior are already large for results from relatively large volumes, compared to the scales set by our choice of parameters. In order to have a sufficiently large set of results to extract the scaling behavior, we also take results with non-negligible scaling corrections into account.

As outlined above, including the first explicitly volume-dependent correction term, the order parameter can be expanded as
\be
M(z, h, L)&=& L^{-\beta/\nu} \left[ Q_M(z, h L^{ \beta \delta/\nu}) + \frac{1}{L^\omega} Q^{(1)}_M(z, h L^{\beta \delta/\nu}) + \ldots \right].
\label{eq:Mfsscorr}
\ee
In order to extract the scaling functions for the order parameter, we start from the assumption that the deviation from the perfect scaling behavior contained in the scaling function $Q_M(z, \bar{h})$ is completely determined by the function $Q^{(1)}_M$ in Eq.~\eqref{eq:Mfsscorr}. We will confirm that this assumption is sufficiently well satisfied by our results.

We can remove the leading-order scaling function by comparing the rescaled order parameter for different volume size, but at the same value of the scaling variables $z$ and $\bar{h} = h L^{\beta \delta/\nu}$. This determines the coefficient function $Q_M^{(1)}(z, \bar{h})$ of the non-universal correction term. For two different volume sizes $L_1$ and $L_2$, we choose values for the external fields $h_1$ and $h_2$ such that 
\be
\bar{h}^* = \bar{h}_1 =h_1 L_1^{\beta\delta/\nu} =  \bar{h}_2 = h_2 L_2^{\beta\delta/\nu}. 
\ee
Since the finite-size scaling function depends only on $z$ and $\bar{h}$, the universal term $Q_M(z, \bar{h})$ drops out in  the difference of the two values and we are left with
\be
L_1^{\beta\delta/\nu} M(z, h_1, L_1) - L_2^{\beta\delta/\nu} M(z, h_2, L_2) &=& \frac{1}{L_1^\omega} Q^{(1)}_M(z, h_1 L_1^{\beta\delta/\nu})  - \frac{1}{L_2^\omega} Q^{(1)}_M(z, h_2 L_2^{\beta\delta/\nu}) + \ldots  \nn\\
&=&  Q_M^{(1)}(z, \bar{h}^*)  \frac{1}{L_2^\omega}  \left( s^\omega  -1 \right) + \ldots
\label{eq:scalingcorrections}
\ee 
with $s= L_2/L_1$. By fitting this expression as a function of the size ratio $s$ to results for different volume sizes, the coefficient function as well as the exponent $\omega$ can be determined.

Scaling corrections due to large values of the external symmetry-breaking field and the small volume are 
both present in our results. 
Since these scaling corrections become progressively larger the further a way one moves from the critical temperature and the smaller the volume, we have restricted the results for large values of the variable $z$ to larger volume sizes. For $|z| \le 1.0$, we take volume sizes from $L = 100$ fm down to $L = 1 $ fm  into account, while for $|z| > 1.0$ we only use the volumes from $L= 100$ fm down to $L = 10$ fm to extract the finite-size scaling functions. 
\section{Results and Discussion}
\label{sec:results}
In this section, we present the results from our investigation of the finite-size scaling behavior of the O(4) model in $d=3$ dimensions and from our determination of the finite-size scaling functions.
We start with a discussion of the finite-size scaling behavior of the order parameter, illustrate the extraction of the scaling functions, and propose a parameterization for these functions for the limits of small and large values of the scaling variable $\bar{h}$. We continue with an analogous discussion of the finite-size scaling behavior of the susceptibility.  
We finally show results for the finite-size scaling functions $Q_M(z, \bar{h})$ and $Q_\chi(z, \bar{h})$ for a selected set of values for the scaling variable $z$ as a function of $\bar{h}$ and discuss effects of our approximation scheme.
In the last part of this section, we briefly discuss the scaling functions as functions of the dimensionless correlation length $\xi/L$.

\subsection{Finite-size scaling function for the order parameter}\label{subsec:FSSF_order}
%
%
\begin{figure}
\begin{center}
\includegraphics[scale=0.40, clip=true]{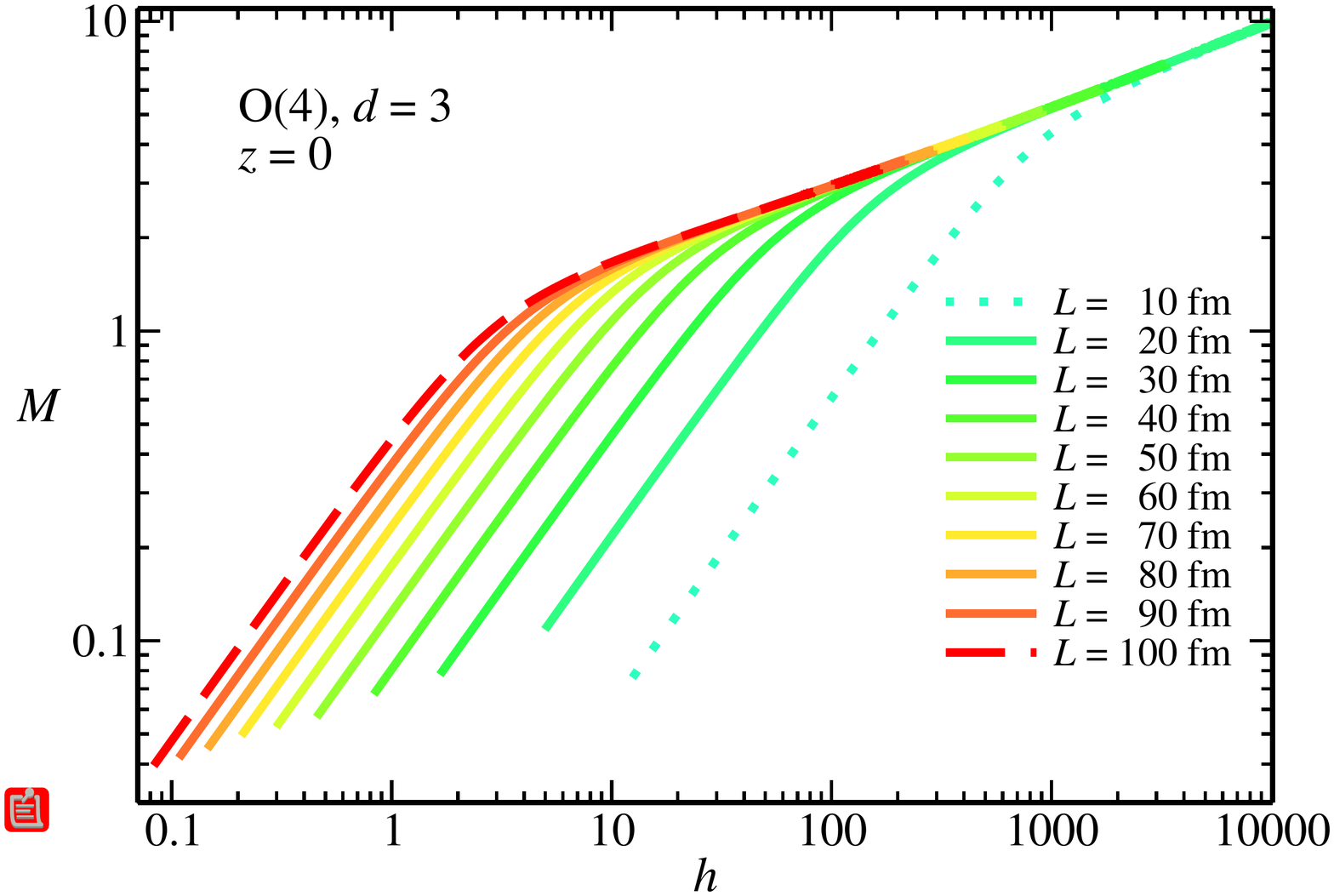}
\includegraphics[scale=0.40, clip=true]{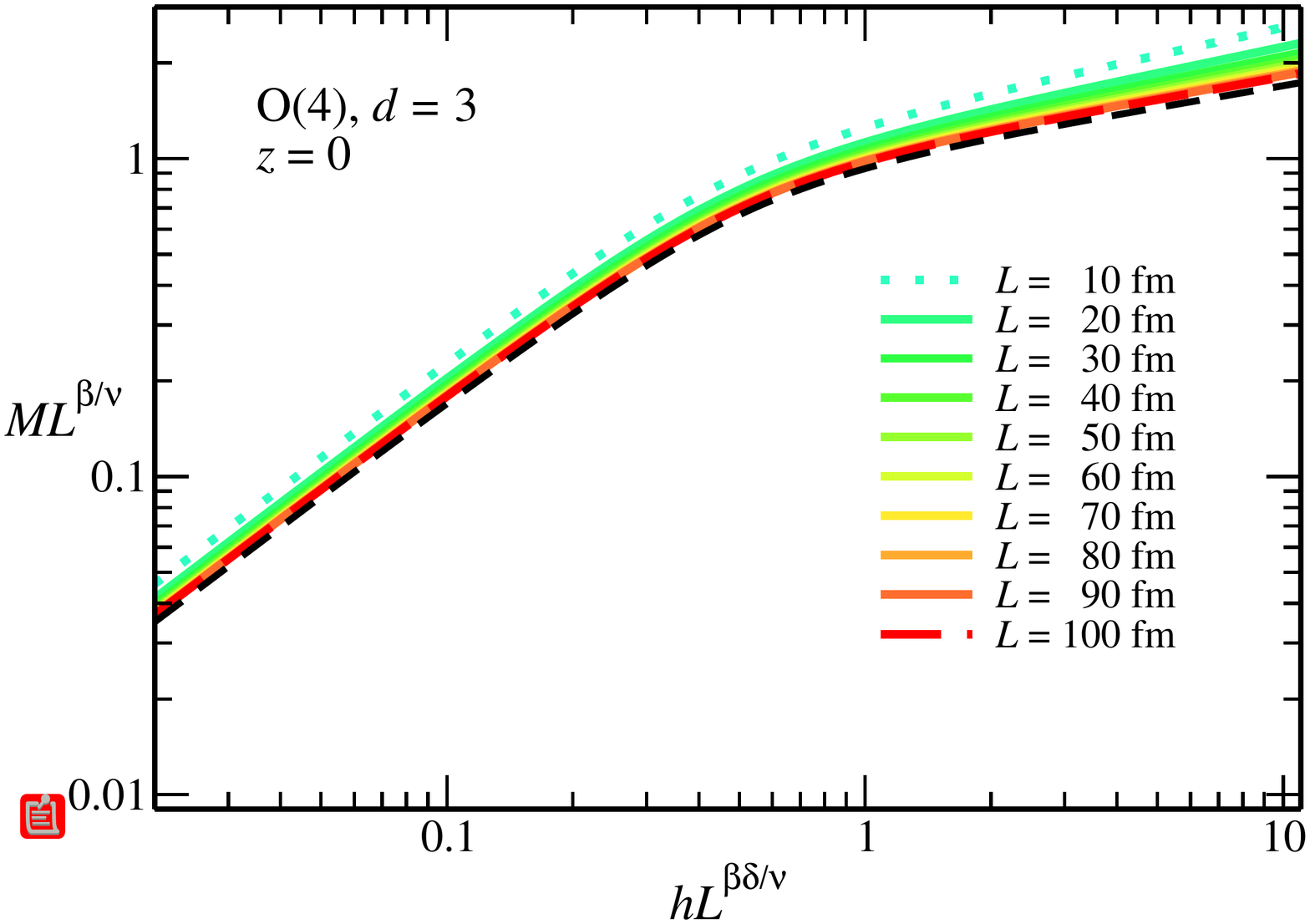}
\end{center}
\caption{Finite-size scaling behavior of the order parameter $M$ at the critical temperature, $z=0$. Shown is the unscaled result for $M$ as a function of $h$ for different volume sizes (first panel), and the finite-size scaled order parameter $L^{\beta/\nu} M$ as a function of  $h L^{\beta \delta/\nu}$ for the same values of the volume size (second panel).
For large values of $h$, where the correlation length is small, the unscaled results for different volume sizes all converge towards the same infinite-volume limit (first panel). For small volume sizes, the corrections to the ideal scaling behavior become considerable.The black dashed line (second panel) is the result for the scaling function obtained from all results by removing the scaling corrections.}
\label{fig:z0orderFVscaled}
\end{figure}
%
%
\begin{figure}
\begin{center}
\includegraphics[scale=0.40, clip=true]{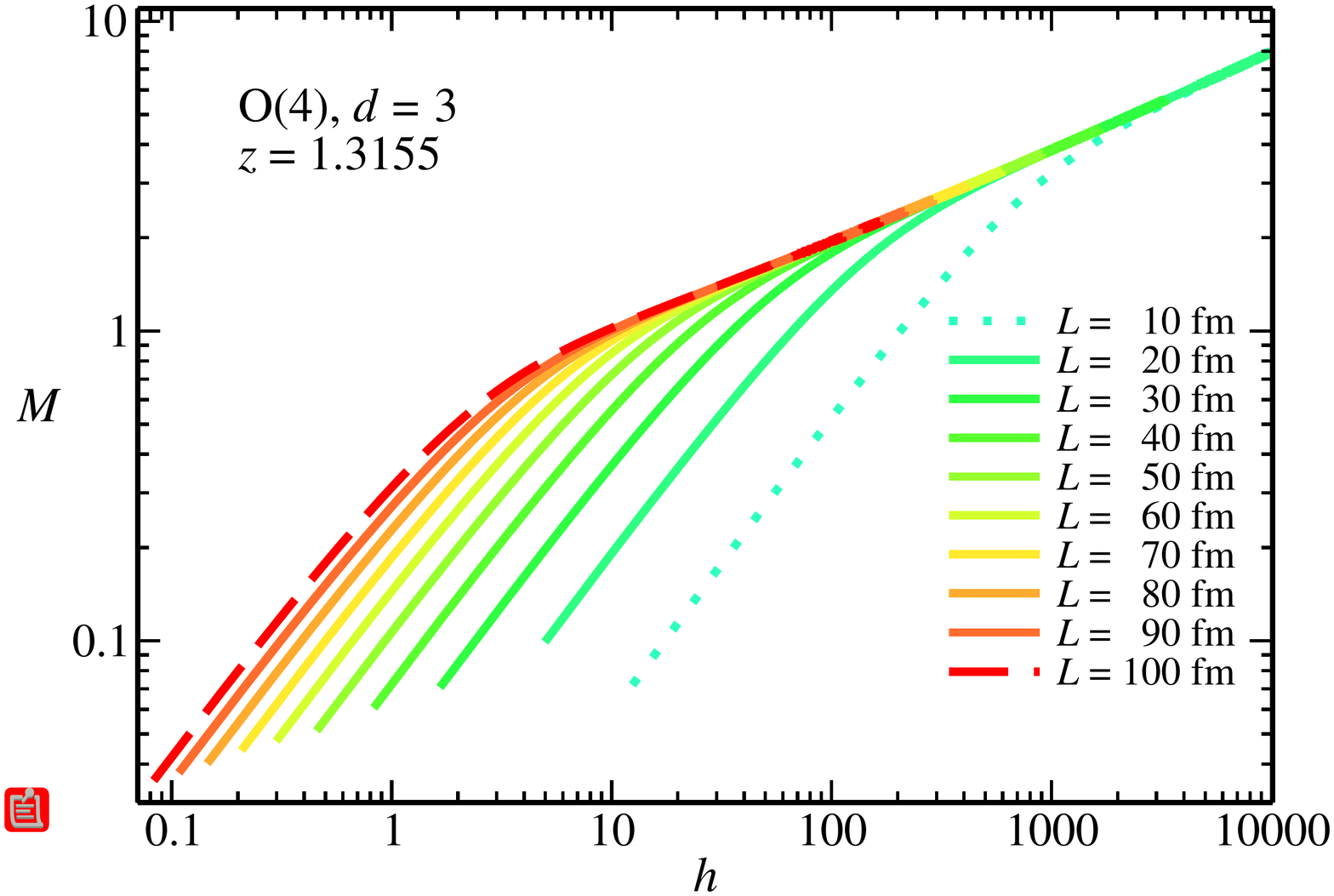}
\includegraphics[scale=0.40, clip=true]{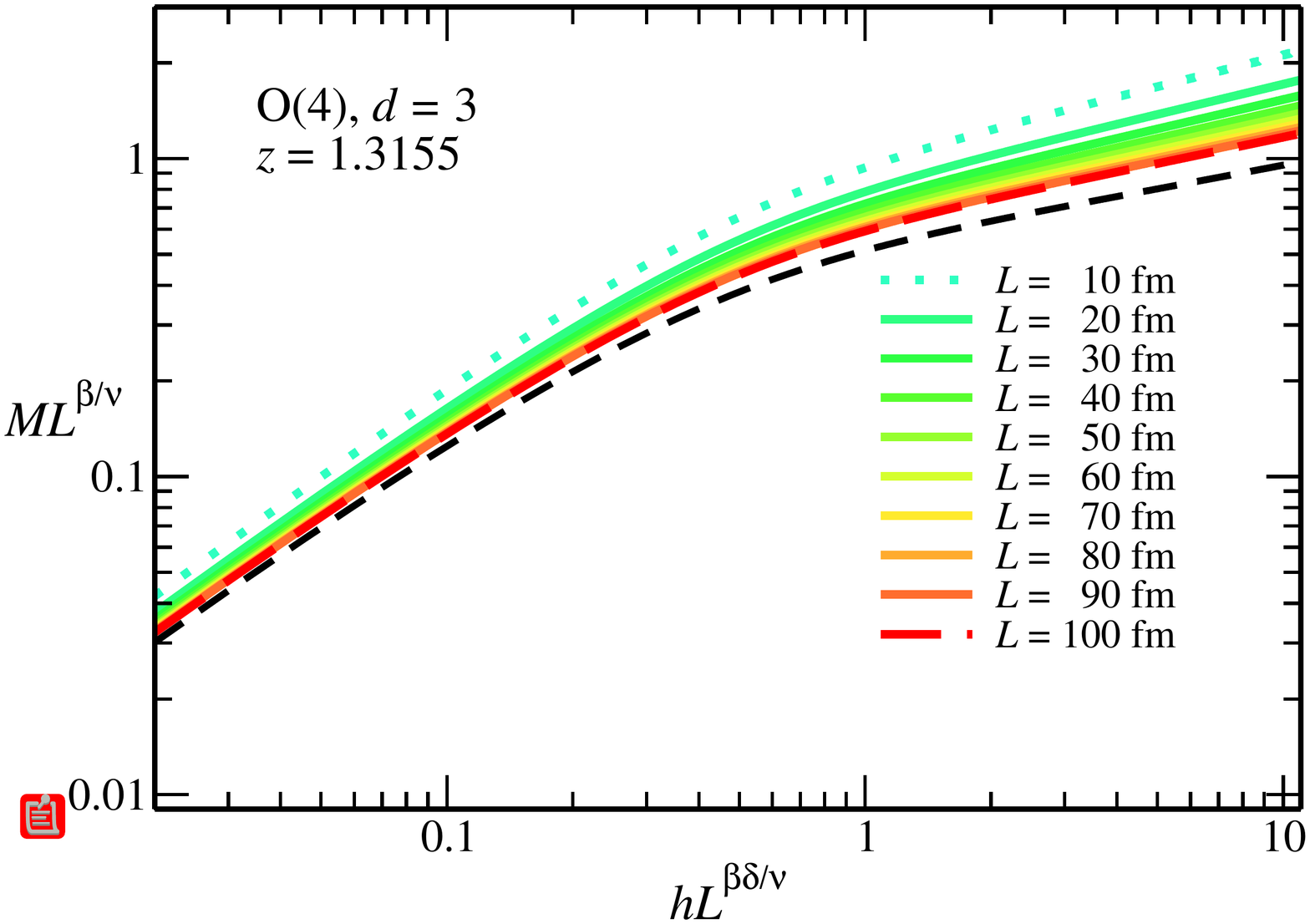}
\end{center}
\caption{Finite-size scaling behavior of the order parameter $M$ at the peak of the longitudinal susceptibility, $z=z_p=1.3155$. Shown is the unscaled result for $M$ as a function of $h$ for different volume sizes (first panel), and the finite-size scaled order parameter $L^{\beta/\nu} M$ as a function of  $h L^{\beta \delta/\nu}$ for the same values of the volume size (second panel).
The black dashed line (second panel) is the result for the scaling function obtained from all results by removing the scaling corrections.}
\label{fig:zpeakorderFVscaled}
\end{figure}
The behavior of the order parameter as a function of the external symmetry-breaking field $h$ for different volume sizes is shown in the first panels of Fig.~\ref{fig:z0orderFVscaled} for 
$z=0$, i. e. at the critical temperature, and Fig.~\ref{fig:zpeakorderFVscaled} for $z=z_p$, i. e. at the peak value of the susceptibility.

Two separate regimes can be distinguished in the double-logarithmic plots: For large values of $h$, the masses of the fluctuations are large and hence the correlation length is much smaller than the volume size. Even the results for small volume size approach asymptotically the infinite-volume limit. As expected from the infinite-volume scaling behavior, the order parameter behaves as $\sim h^{1/\delta}$ and the slope of the curves for large fields is determined by the value of the critical exponent $\delta$. 
For small values of $h$, where the correlation length becomes large, the deviations from the infinite-volume behavior are equally clear: For small volumes, a small correlation length can be of the order of the volume size and the deviations occur already at large values of $h$, whereas for large volumes the correlation length needs to be large for significant finite-size effects which appear only for small values of $h$. 

In the second panels in both Figs.~\ref{fig:z0orderFVscaled} and \ref{fig:zpeakorderFVscaled}, the rescaled order parameter $ML^{\beta/\nu}$ is plotted as a function of the scaling variable $hL^{\beta\delta/\nu}$. 
In the finite-size scaling region, the curves for different values of the volume size collapse almost perfectly onto a single curve. The agreement becomes worse with decreasing volume size, which can be explained by the presence of scaling corrections. When we assume that these scaling corrections can be described by the corrections due to only the first irrelevant operator, we can use the expression from Eq.~\eqref{eq:scalingcorrections} to fit the results. After subtracting the corrections, we obtain a result for the universal leading-order scaling function $Q_M(z, \bar{h})$ (shown as a black dashed line in the figures).

From an RG determination at the fixed point  \cite{Litim:2001hk, Litim:2002cf, Bervillier:2007rc} we expect for the exponent $\omega$ associated with the corrections the value
\be
\omega = 0.7338.
\label{eq:Litimomega}
\ee
For $z=0$, we find from a fit to our rescaled results for $L= 10$ fm to $L=100$ fm for the order parameter in the finite-size scaling region with $\bar{h}\le 1.0$ the value
\be
\omega = 0.7443(300),
\label{eq:orderparameteromega}
\ee
which is compatible with the fixed-point determination. Since we fit to a numerical result that also includes contributions from additional irrelevant operators, our result for $\omega$ should be regarded as an effective value. Having established the consistency of the scaling corrections in our results with the RG prediction, we use in the following the more accurate value \eqref{eq:Litimomega} for the evaluation. 

\begin{figure}
\begin{center}
\includegraphics[scale=0.7, clip=false]{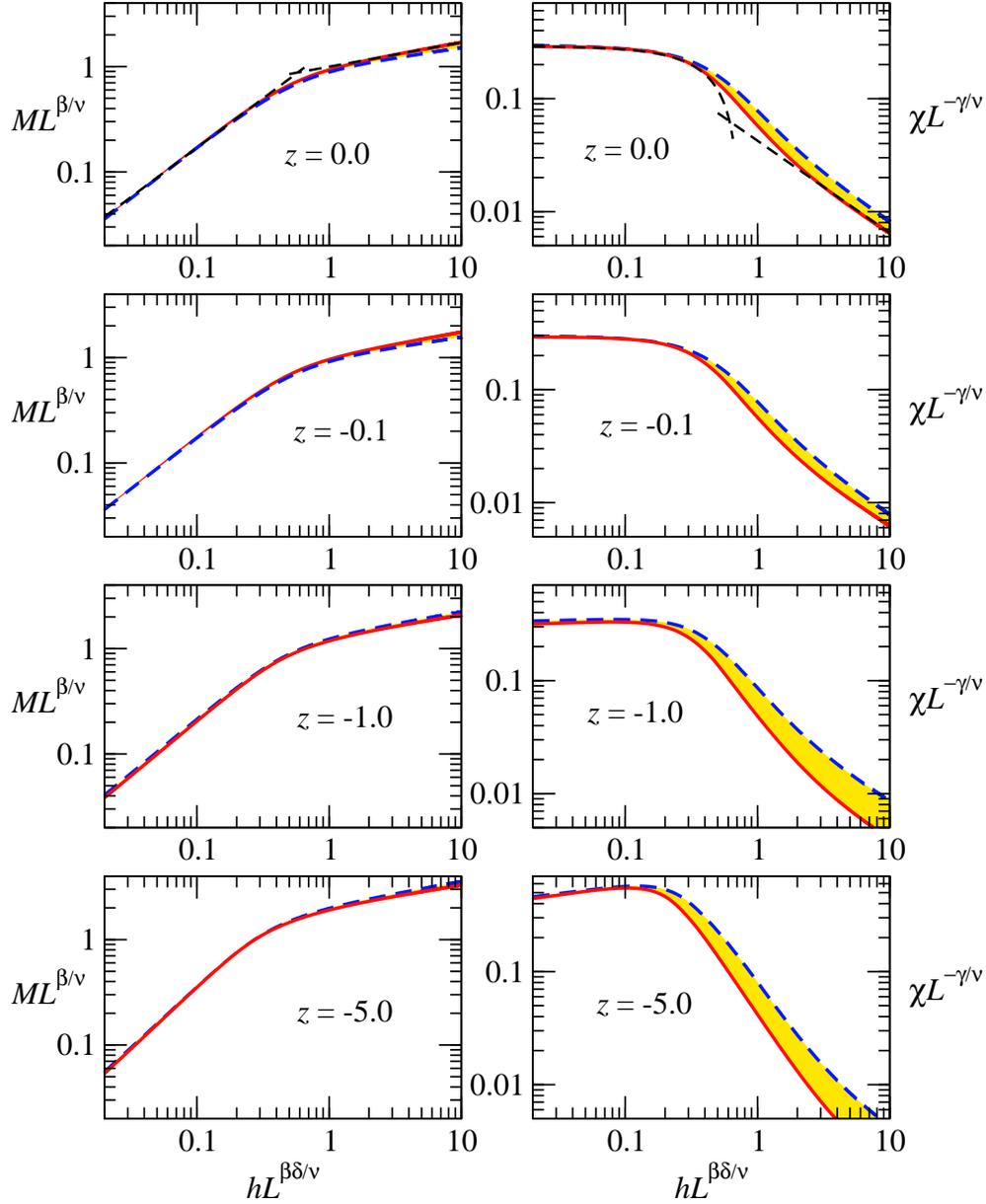}
\end{center}
\caption{Finite-size scaling functions for the order parameter (left column) and the susceptibility (right column) as a function of the finite-size scaling variable $h L^{\beta\delta/\nu}$ for fixed $z \le 0$. Shown are  results obtained from both the PTRG (red solid lines) and the ERG with optimized cutoff (blue dashed lines). Both cutoff schemes agree well in the finite-size scaling region ($h L^{\beta\delta/\nu} < 1$), and deviations appear only for larger values, where the functions are already determined by the infinite-volume behavior. The black dashed lines included for $z=0$ represent the fits to the asymptotic behavior for small and large values of the scaling variable.}
\label{fig:scalingfunctions_z<0}
\end{figure}

We have determined the scaling functions $Q_M(z, hL^{\beta \delta/\nu})$ for a range of $z$-values. Plots for selected values are shown below in Fig.~\ref{fig:scalingfunctions_z>0} for $z\ge 0$, and in Fig.~\ref{fig:scalingfunctions_z<0} for $z<0$. For values $|z| < 1.0$ we have used results from volume sizes $L=1$ fm to $L=10$ fm in steps of $1$ fm and from $L=10$ to $L=100$ fm in steps of $10$ fm. For values $|z|>1.0$, we have only included results from volume sizes from $L=10$ to $L=100$ fm in steps of $10$ fm. 
For small volumes, the value of $h$ has to be increased too much in order to access the finite-size scaling region and scaling corrections become too large.

We have not found a global parameterization for the scaling function, but we find that 
for small values of $h L^{\beta\delta/\nu}$ our results are fitted well by the parameterization 
\be
Q_M^{(0)}(z, hL^{\beta \delta/\nu}) = c(z) (h L^{\beta\delta/\nu})^{\tau(z)}.
\ee
Both the coefficient $c(z)$ and the exponent $\tau(z)$ vary with $z$. The values we determined for the fitting parameters can be found in Tab.~\ref{tab:QMparasmallhsmallz} and Tab.~\ref{tab:QMparasmallhlargez} for $|z|<1.0$ and $|z|\ge 1.0$, respectively. We include in the tables the approximate values for $hL^{\beta\delta/\nu}$ below which the fit is applicable. For comparison to the actual scaling functions, we also show the fits in the first panels of the Figs.~\ref{fig:scalingfunctions_z>0} and \ref{fig:scalingfunctions_z<0}.

For large values of $h L^{\beta\delta/\nu}$, the parameterization is constrained by the asymptotic behavior of the scaling functions for infinite volume:
\be
Q_M^{(0)}(z, hL^{\beta \delta/\nu}) = c_\infty(z) (h L^{\beta\delta/\nu})^{\tau_\infty}.
\ee
We expect the exponent $\tau_\infty$ to be independent of $z$ and to have the value $\tau_\infty = 1/\delta= 0.2011(1)$ for our values for the critical exponents. The coefficient $c_\infty(z)$ retains its $z$-dependence and is expected to coincide with the value of the infinite-volume scaling function, $c_\infty(z) = f(z)$. 
The results from the fits are summarized in Tab.~\ref{tab:QMparaasympsmallz} for $|z|< 1.0$ and Tab.~\ref{tab:QMparaasymplargez} for $|z|\ge 1.0$. For comparison with $c(z)$ and $\tau_\infty$, the corresponding values for $f(z)$ are included in the tables. The values for the coefficient $c_\infty(z)$ and $f(z)$ agree within $1\%$, whereas the agreement of the exponent $\tau_\infty$ with the expected value is only fair.
%
%
\begin{figure}
\includegraphics[scale=0.7, clip=true]{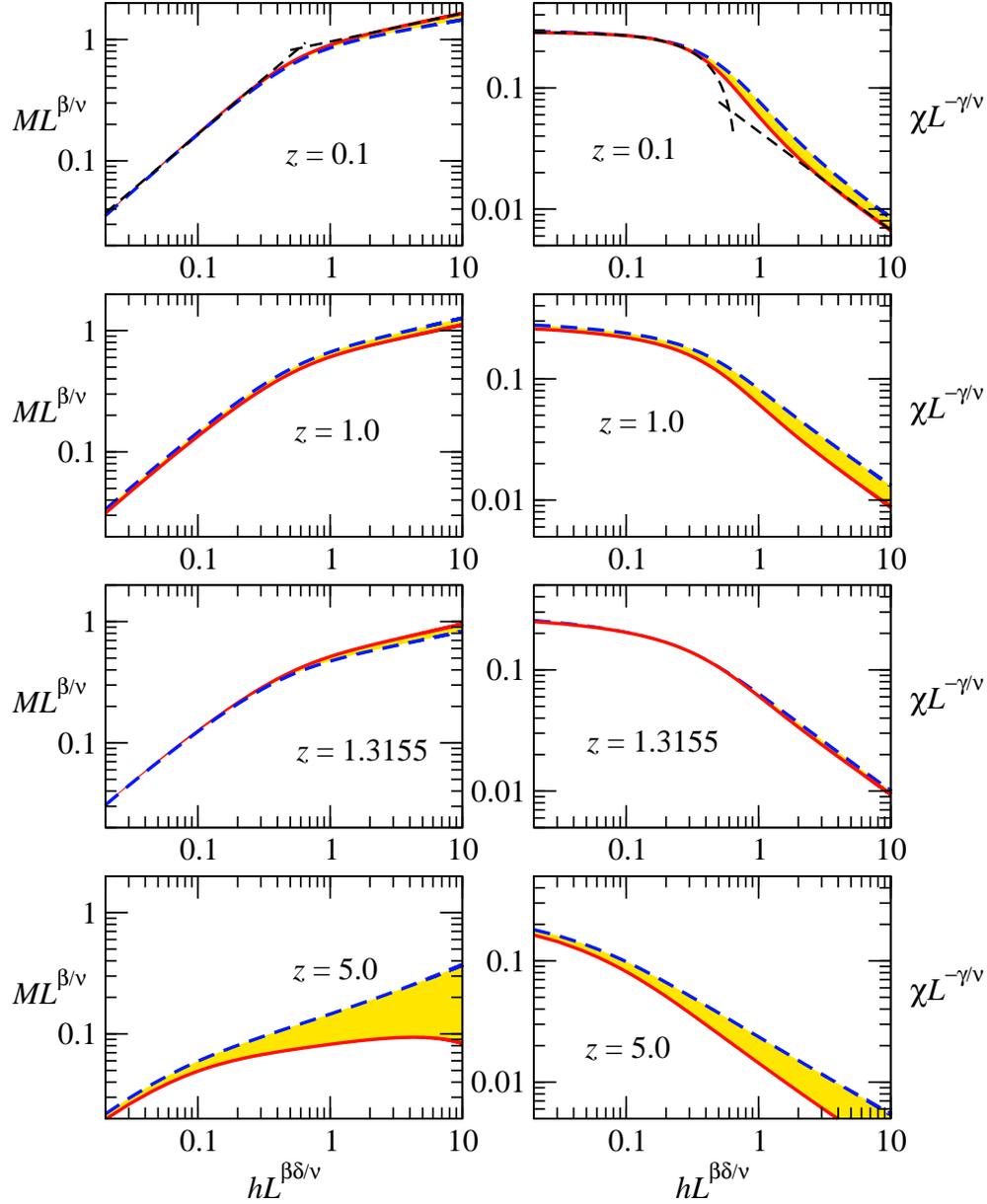}
\caption{Finite-size scaling functions for the order parameter (left column) and the susceptibility (right column) as a function of  $h L^{\beta\delta/\nu}$ for fixed  $z > 0$ from PTRG (red solid line) and ERG with optimized cutoff (blue dashed line). For $z=0.1$, the results of the fits to the asymptotic behavior for small and large values of the scaling variable are included (black dashed lines). The asymptotic behavior is described well, but the parameterization is clearly not applicable in the intermediate region.}
\label{fig:scalingfunctions_z>0}
\end{figure}
%
%
\clearpage

\subsection{Finite-size scaling function for the susceptibility}
\begin{figure}[t]
\begin{center}
\includegraphics[scale=0.40, clip=true]{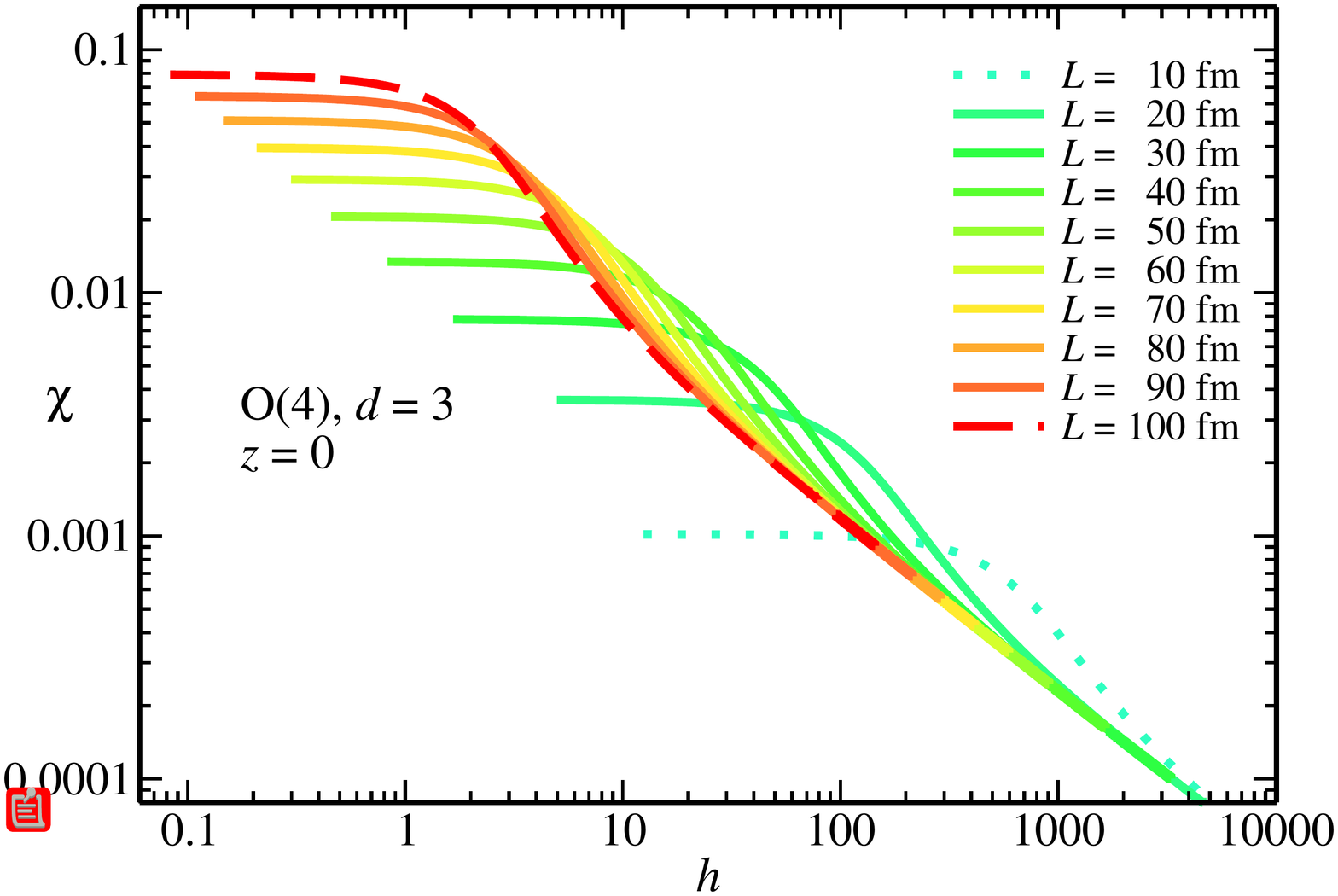}
\includegraphics[scale=0.40, clip=true]{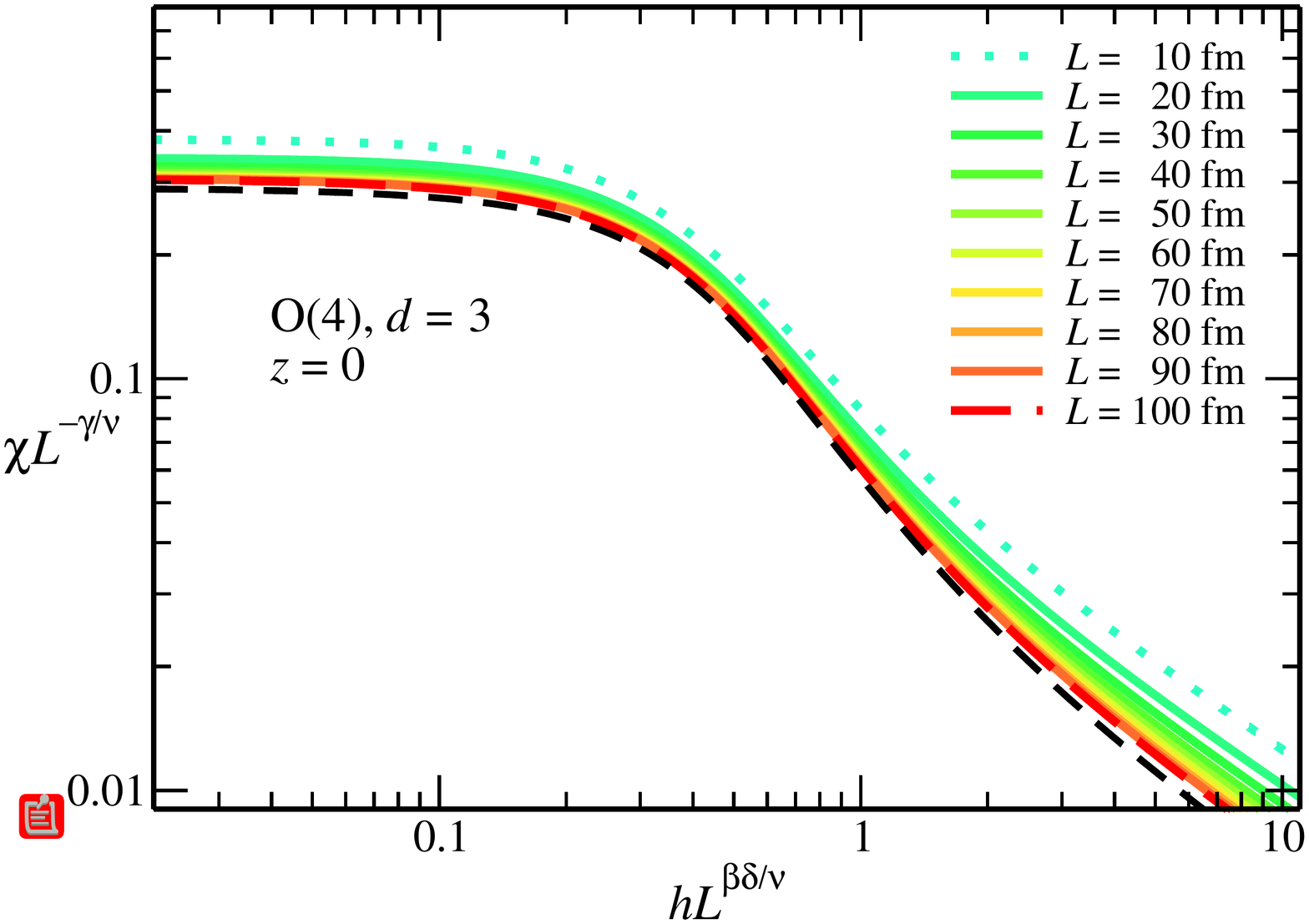}
\end{center}
\caption{Finite-size scaling behavior of the longitudinal susceptibility at the critical temperature, $z=0$. The susceptibility for different volume sizes is shown as a function of $h$ (first panel). For large values of $h$, the results for different volume sizes all converge towards the infinite-volume limit. The finite-size scaled susceptibility  $L^{-\gamma/\nu}\chi$ for the same volume sizes is shown as a function of $h L^{\beta \delta/\nu}$ (second panel). For small volume sizes, the corrections to the ideal scaling behavior are of considerable size. The result for the scaling function obtained from all results by removing the scaling corrections is shown as a black dashed line.}
\label{fig:suscscalingz0}
\end{figure}

\begin{figure}[t]
\begin{center}
\includegraphics[scale=0.40, clip=true]{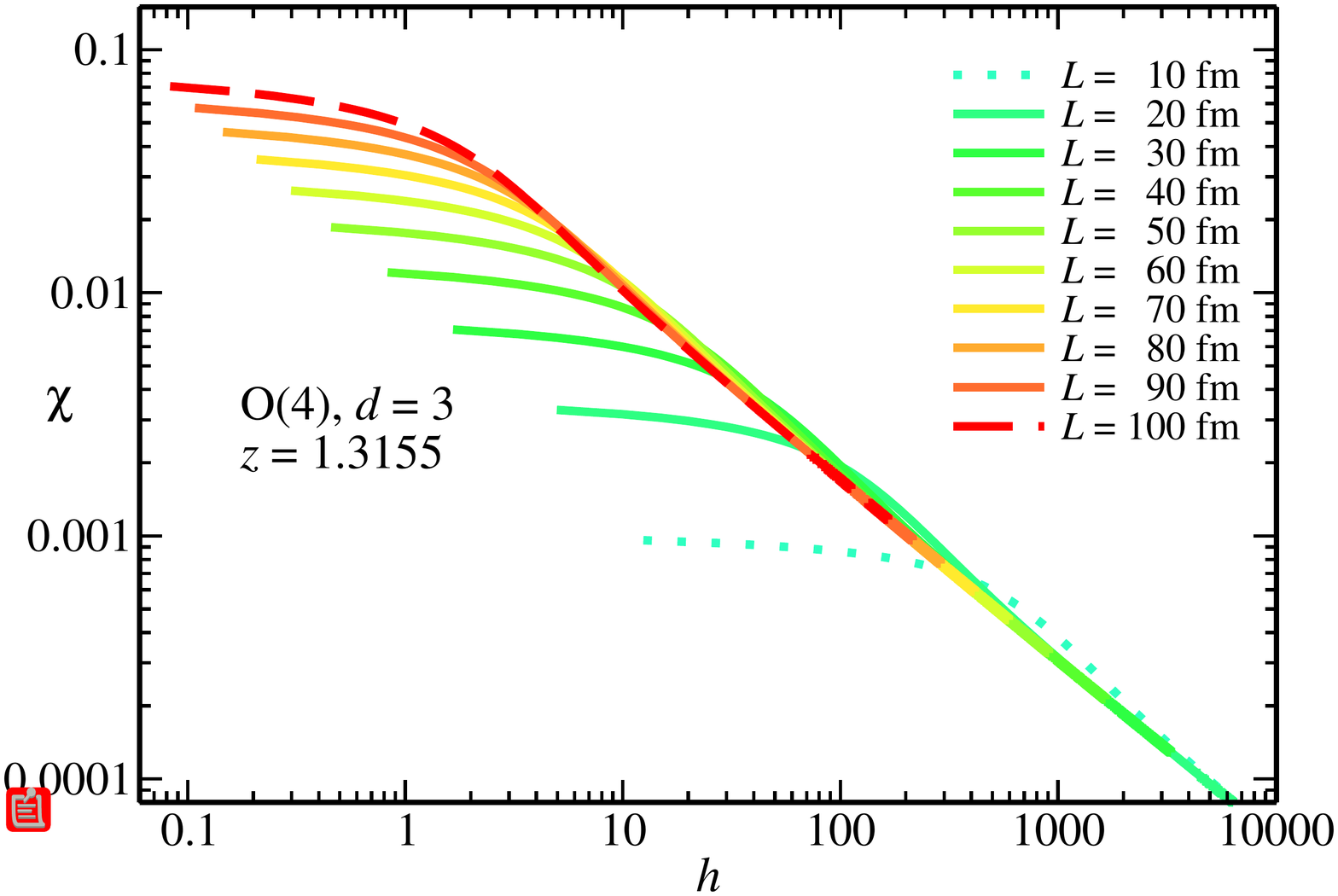}
\includegraphics[scale=0.40, clip=true]{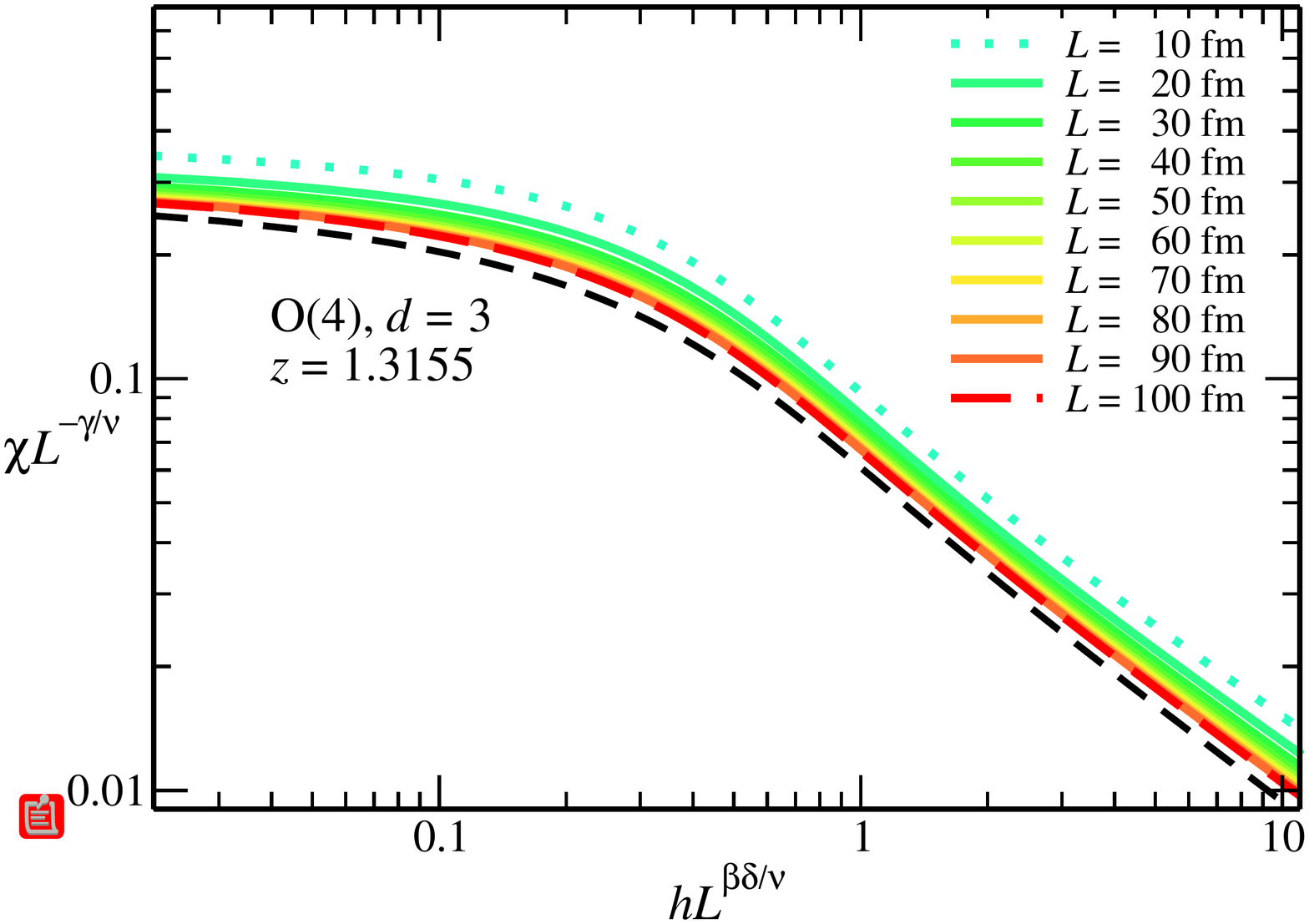}
\end{center}
\caption{Finite-size scaled longitudinal susceptibility at the peak value, $z=z_p=1.3155$. The susceptibility $\chi$ for different volume sizes is shown as a function of $h$ (first panel). For large values of $h$, where the correlation length is small, the results for different volume sizes all converge towards the same infinite-volume limit. The finite-size scaled susceptibility $L^{-\gamma/\nu}\chi$ is shown as a function of $h L^{\beta \delta/\nu}$ (second panel). Corrections to the scaling behavior can be seen for the small volume sizes. The scaling function obtained by removing the scaling corrections is represented by the black dashed line.}
\label{fig:suscscalingzpeak}
\end{figure}

The scaling behavior of the susceptibility $\chi$ as a function of the field $h$ is shown in Figs.~\ref{fig:suscscalingz0} and \ref{fig:suscscalingzpeak} for $z=0$ and $z=z_p=1.3155$, respectively. We can once again distinguish an asymptotic region for large $h$, where the correlation length is much smaller than the volume size and the results converge towards the infinite-volume limit, behaving as $\sim h^{1/\delta-1}$. For small values of $h$, the finite-size effects are strong and the susceptibility tends towards a constant value which depends on the volume and varies by about two orders of magnitude when the volume is changed by a factor of $10$.
%
%
%

In both Figs.~\ref{fig:suscscalingz0} and \ref{fig:suscscalingzpeak} the results for the rescaled susceptibility $L^{-\gamma/\nu} \chi$ are shown in the second panels as a function of the scaling variable $hL^{\beta \delta/\nu}$. As for the order parameter, the collapse of the rescaled results onto the scaling function is quite good, although not perfect due to scaling corrections for smaller volume size. We use once again the RG description of these scaling corrections to extract the leading-order scaling function, which is depicted by a dashed black line in the figures.

Scaling corrections from the results for the rescaled susceptibility for $z=0$ at $\bar{h}\le 1.0$ lead to a value of the exponent 
\be
\omega = 0.7384(400)
\ee
which is in agreement with the result Eq.~\eqref{eq:orderparameteromega} from the order parameter, and with the result Eq.~\eqref{eq:Litimomega} from the fixed point determination \cite{Litim:2001hk, Litim:2002cf, Bervillier:2007rc}. As for the order parameter, we use the more accurate result \eqref{eq:Litimomega} for all further evaluations.

The determination of the finite-size scaling function and the values included in the evaluation are the same as outlined above for the order parameter.

As for the order parameter, we have not found a satisfactory global parameterization for the scaling function $Q_\chi(z, \bar{h})$. However, we are able to parameterize the functions for small and large values of $\bar{h}$ separately.

For small values of $\bar{h}$, we find that the parameterization $ d(z) + c(z) (hL^{\beta \delta/\nu})^\tau$ works sufficiently well in the range of $z$-values that we consider.
The parameterization of the scaling function for the susceptibility is in principle constrained by that of the order parameter. 
Because the exponent $\tau$ of the parameterization for the order parameter is close to one, $\tau-1\ll 1$, the leading term of the derivative of the scaling function $Q_M(z, \bar{h})$ with respect to $h$ is approximately constant, which is consistent with the ansatz for the parameterization that we have chosen here for the susceptibility. Beyond that, the compatibility of the results for susceptibility and order parameter cannot be checked very well using this parameterization. We will discuss this in more detail below and demonstrate for the special case $z=0$ that our results for the order parameter and the susceptibility are indeed consistent. 
The values for the fit parameters determined for small values of $\bar{h}$ are given in Tab.~\ref{tab:suscparasmallhsmallz} for small values of $|z|<1.0$ and in Tab.~\ref{tab:suscparasmallhlargez} for values of $|z|\ge1.0$. The last column in the table lists the approximate values below which the parameterization is valid. For comparison to the scaling functions, the fits are shown in the first panels of  Figs.~\ref{fig:scalingfunctions_z>0} and \ref{fig:scalingfunctions_z<0}.

For large values of the scaling variable, the parameterization of the scaling function is once again constrained by the known behavior in the infinite-volume limit. We use the parameterization
$c_\infty(z) (hL^{\beta \delta/\nu})^{\tau_\infty}$ for large values of $\bar{h}$.
We expect that the exponent is independent of $z$ and takes the value $\tau_\infty = 1/\delta -1=-0.7989(1)$ with our value for $\delta$.
For the coefficient, we expect that it coincides with the infinite-volume scaling function for the susceptibility, $c_\infty(z) = \frac{1}{H_0}f_\chi(z)$. The results of the fits are summarized in Tabs.~\ref{tab:suscparaasympsmallz} and \ref{tab:suscparaasymplargez} for $|z|<1.0$ and $|z|\ge 1.0$, respectively. For comparison, values for $H_0 c_\infty(z)$ and $f_\chi(z)$ are included in the same tables. 
The agreement for the coefficient $c_\infty(z)$ with the scaling function $f_\chi(z)$ is not as good as the agreement in the corresponding relation for the order parameter, although still quite good
close to the critical temperature ($z=0$) and the susceptibility peak ($z=z_p=1.3155$). The agreement of the exponent $\tau_\infty$ with the expectation is good for small values 
of $z$, but for both parameters the agreement becomes progressively worse when one moves to large absolute values of $z$.

%
\begin{figure}
\begin{center}
\includegraphics[scale=0.60, clip=true]{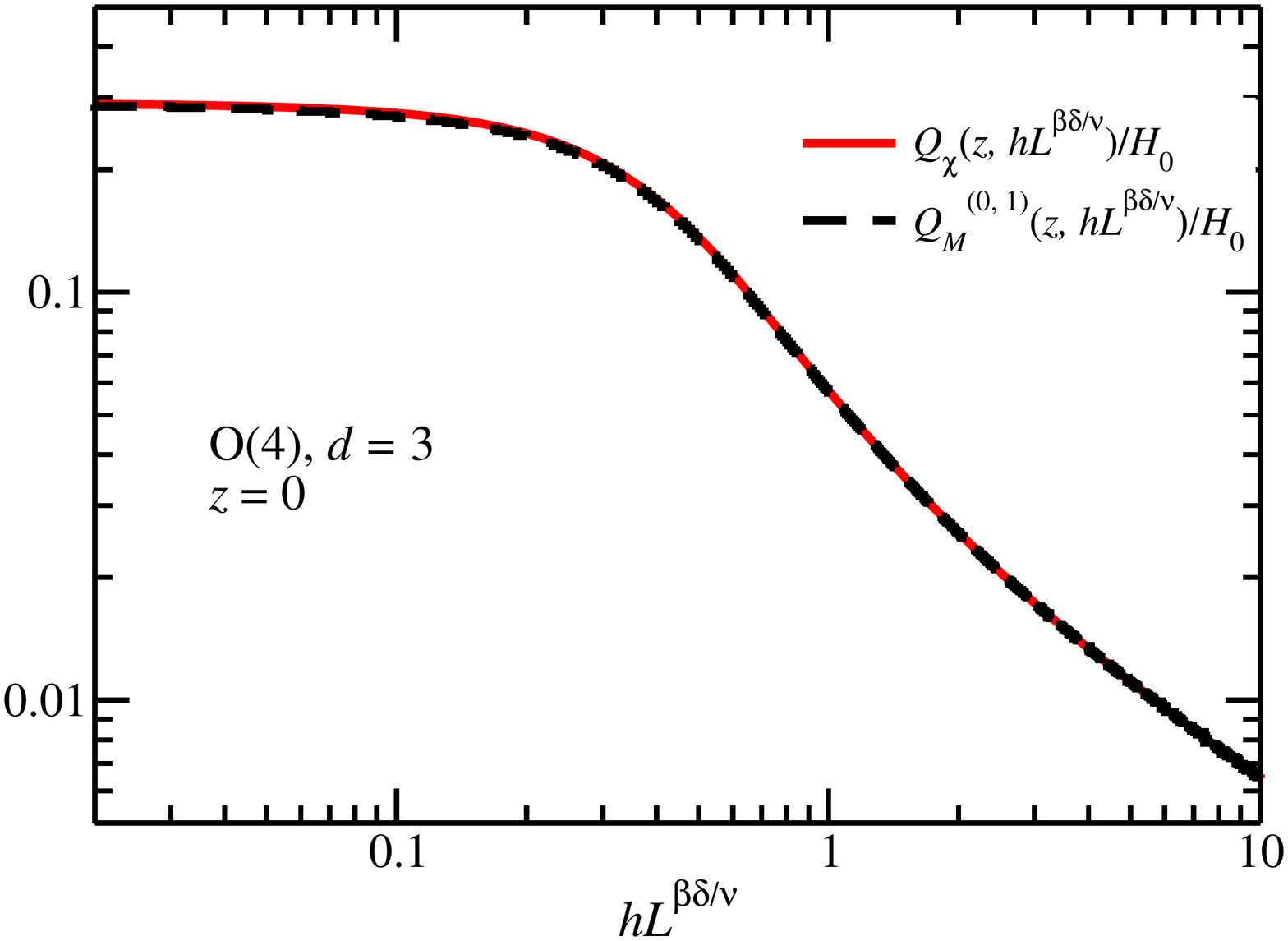}
\end{center}
\caption{Comparison of the finite-size scaling function $Q_\chi(z, \bar{h})$ determined from the susceptibility (red solid line) and  $Q_{M}^{(0, 1)}(z, \bar{h})$, the derivative of the scaling function for the order parameter with respect to  $\bar{h}$ (black dashed line), for $z=0$, as obtained from the PTRG. 
As expected, they agree well and are almost indistinguishable on the scale of the figure. This provides strong evidence that the truncation of the RG flow is sufficient to account for the relevant scaling behavior.}
\label{fig:z0susccheck}
\end{figure}

As discussed in Sec.~\ref{sec:FS_scaling}, the finite-size scaling functions $Q_M(z, \bar{h})$ and $Q_\chi(z, \bar{h})$ of the order parameter and the susceptibility are connected, and in fact the scaling function for the susceptibility should be completely specified by the one for the order parameter. Due to the dependence of the fit parameters on the scaling variable $z$ and the limited number of $z$-values for which we determined these parameters, we cannot test this relation in terms of the parameterization. In addition, because of the additional $z$-dependence we also cannot test the relation numerically for arbitrary value for $z$. However, as we have seen from Eq.~\eqref{eq:suscrelation}, for the special case $z=0$ additional $z$-dependent terms drop out and we can check the validity of this relation.
The comparison between the actual scaling function $Q_\chi(z, \bar h)$ obtained from the susceptibility and the derivative of the scaling function $Q_M(z, \bar h)$ is shown 
in Fig.~\ref{fig:z0susccheck} as obtained from our PTRG approach.
We find that the scaling function obtained from the susceptibility (solid red line) and the derivative of the scaling function for the order parameter (dashed black line) agree very
well, and are indistinguishable on the scale of the plot.
This demonstrates that the RG flow equations capture the scaling behavior consistently and that the results do indeed satisfy the expected scaling relations, as we have already seen in the application to infinite volume \cite{Braun:2007td}. With regard to the truncation of the RG flow, the agreement demonstrates that the truncation of the RG flow is sufficiently large for our calculation.

The results for the scaling functions for both the order parameter and the susceptibility shown in Figs.~\ref{fig:scalingfunctions_z>0} and \ref{fig:scalingfunctions_z<0} have been obtained from an RG calculation using a proper-times cutoff (PTRG) (red lines) and from an ERG calculation using an optimized cutoff (blue lines).
For infinite volume, both calculations coincide and the results are exactly the same. In addition to the agreement in infinite volume, the result again coincide for very small values, where all fluctuations are cut off by the finite volume size. In finite volume, the threshold functions that enter into both calculations differ, see the discussion in 
Sec.~\ref{sec:FV_RG}. Due to truncation effects, which affect the finite-volume RG flows in both calculations in slightly different ways, the results do not coincide in the intermediate-volume 
region (where $2.5  \lesssim L/\xi \lesssim  5$). Note that the double-logarithmic scale in Figs.~\ref{fig:scalingfunctions_z>0} and \ref{fig:scalingfunctions_z<0} magnifies the apparent difference between the results for large volume.
For asymptotically large external symmetry breaking fields (mean-field limit), we find agreement of the results from both approaches again. Overall, we observe that  the agreement is better 
for the order parameter than for the susceptibility, and we conclude that the truncation errors affect the higher-order couplings that enter into e.g. the susceptibility more strongly than the order 
parameter. 
The difference between the methods can be used to assess the truncation errors and provides a theoretical error estimate.

\clearpage

\subsection{Scaling results as a function of the correlation length}\label{subsec:SR_corrlength}
As discussed in Sec.~\ref{sec:FS_scaling}, for comparisons of different systems in the same universality class it is advantageous to use dimensionless ratios since it eliminates the need for a determination of non-universal normalization constants. For this reason, we briefly present results for the scaling behavior as a function of the dimensionless finite-volume correlation length $\xi(L)/L$ in this section.

\begin{figure}
\includegraphics[scale=0.5, clip=true]{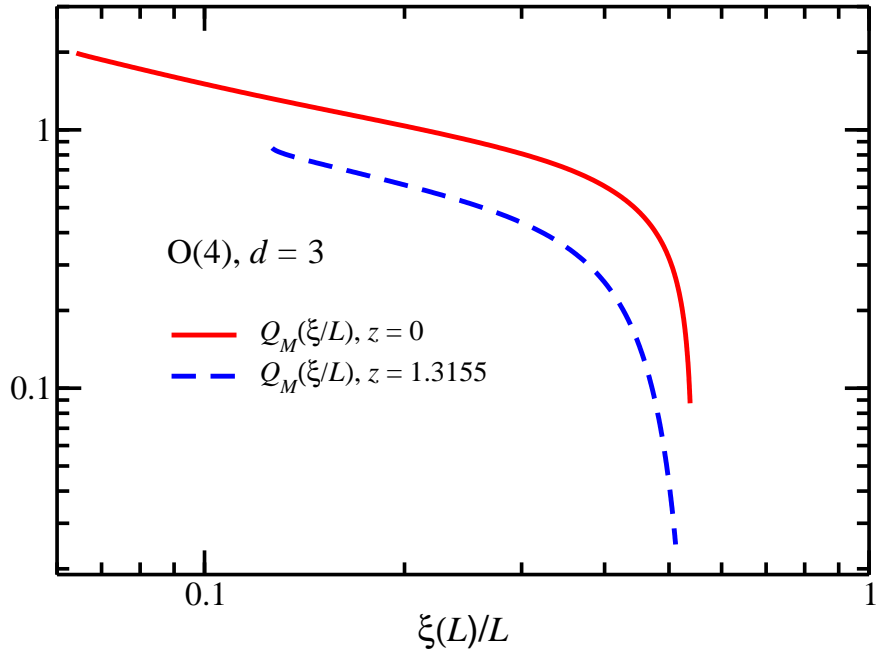}
\caption{Finite-size scaling function $Q_M(\xi/L)$ for the order parameter as a function of the dimensionless finite-volume correlation length, $\xi/L$. The results shown are for the critical temperature ($z=0$, red solid line) and for the peak of the susceptibility ($z=z_p=1.3155$, blue dashed line). Plotting the scaling function as a function of the dimensionless quantity $\xi/L$, we eliminate the need for the determination of an additional non-universal length normalization factor. We find that the finite-volume correlation length is bounded at approximately $\xi/L \approx 0.5$.}
\label{fig:QofxioverL}
\end{figure}

Using the relation $\chi(z, h, L) = \xi^2(z, h, L)$ for $\eta=0$, we translate the scaling function for the order parameter $Q_M(z, \bar{h})$ into a function $Q_M(z, \xi(L)/L)$ of $\xi(L)/L$. 
The results for $z=0$ and $z=z_p$ are shown in Fig.~\ref{fig:QofxioverL}. Since the susceptibility is bounded for a given volume, the correlation length is bounded as well, and we observe that this bound is approximately at $\xi(L)/L \approx 0.5$.
%

\begin{figure}
\includegraphics[scale=0.28, clip=true]{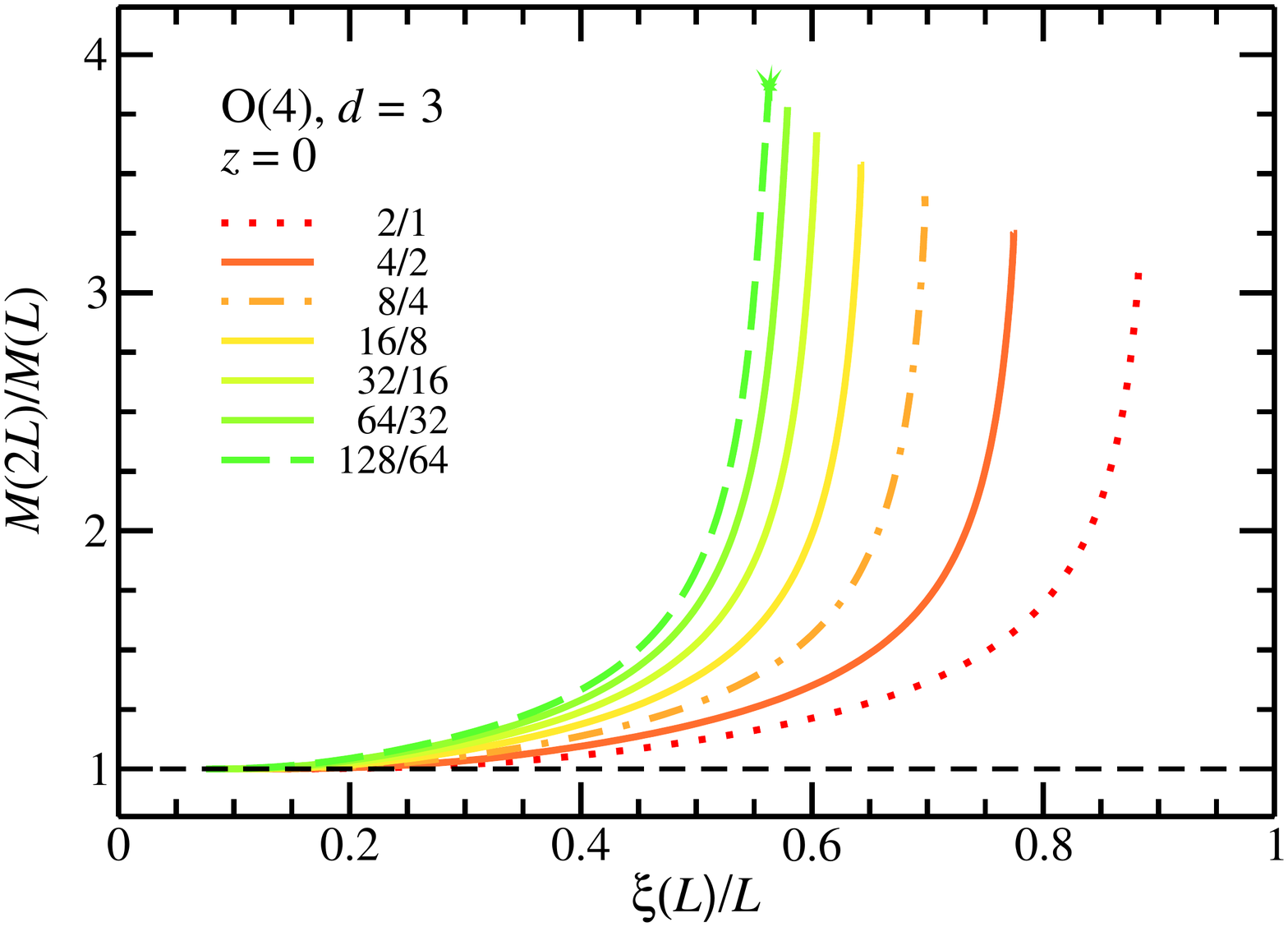}
\includegraphics[scale=0.28, clip=true]{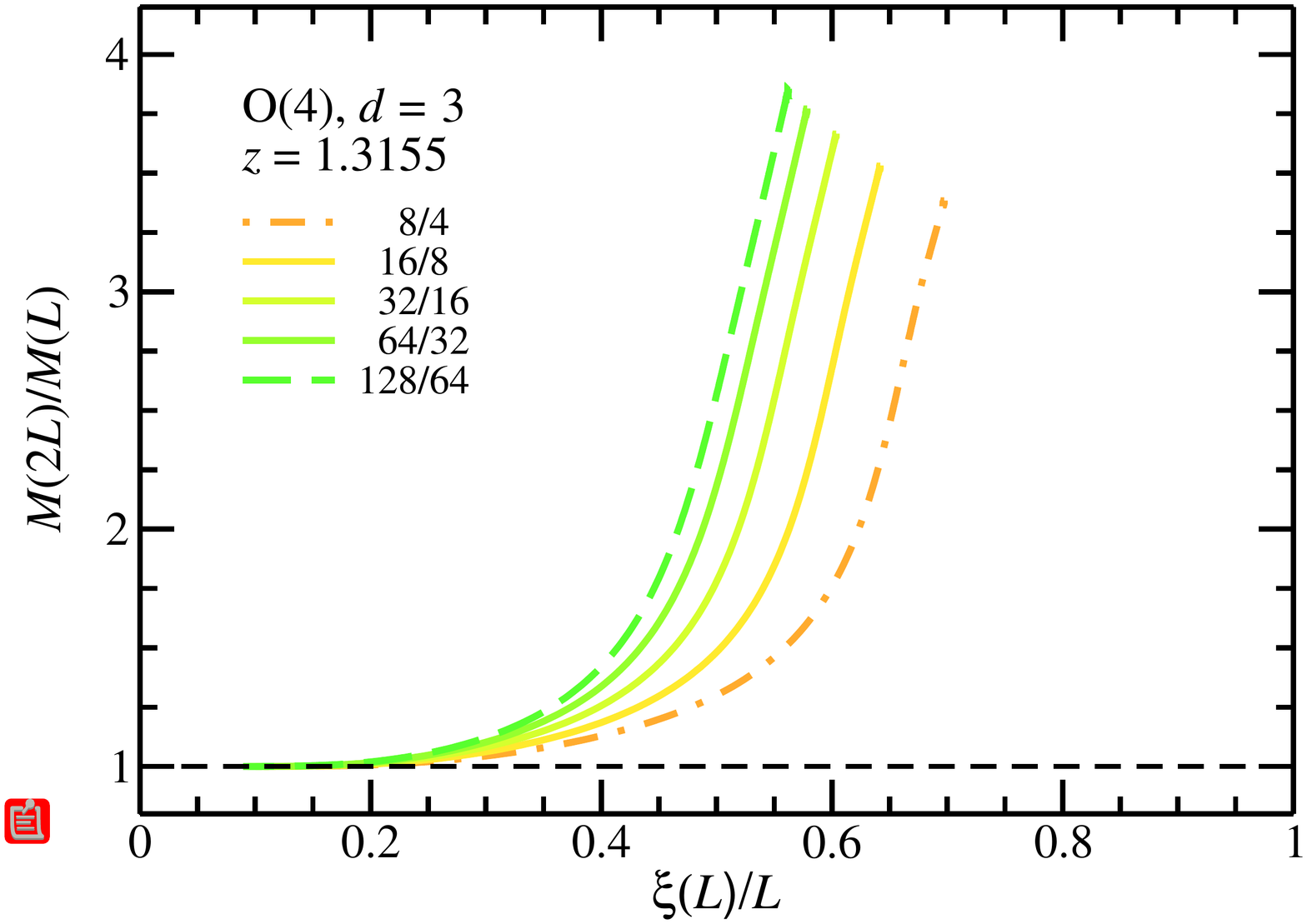}
\caption{Ratio of the order parameter $M(2L)/M(L)$ for systems with the volume sizes $2L$ and $L$, shown as a function of the ratio of the finite-volume correlation length and the system size, $\xi(L)/L$. Here the results are given for $z=0$ (at the infinite-volume value for the critical temperature) (first panel) and $z=z_p=1.3155$ (at the peak value of the susceptibility) (second panel). Corrections to scaling are large for small volume sizes ($1-2$ fm for our choice of parameters), but the results converge for large volume size ($\sim 100$ fm).}
\label{fig:Mratios}
\end{figure}

\begin{figure}
\includegraphics[scale=0.28, clip=true]{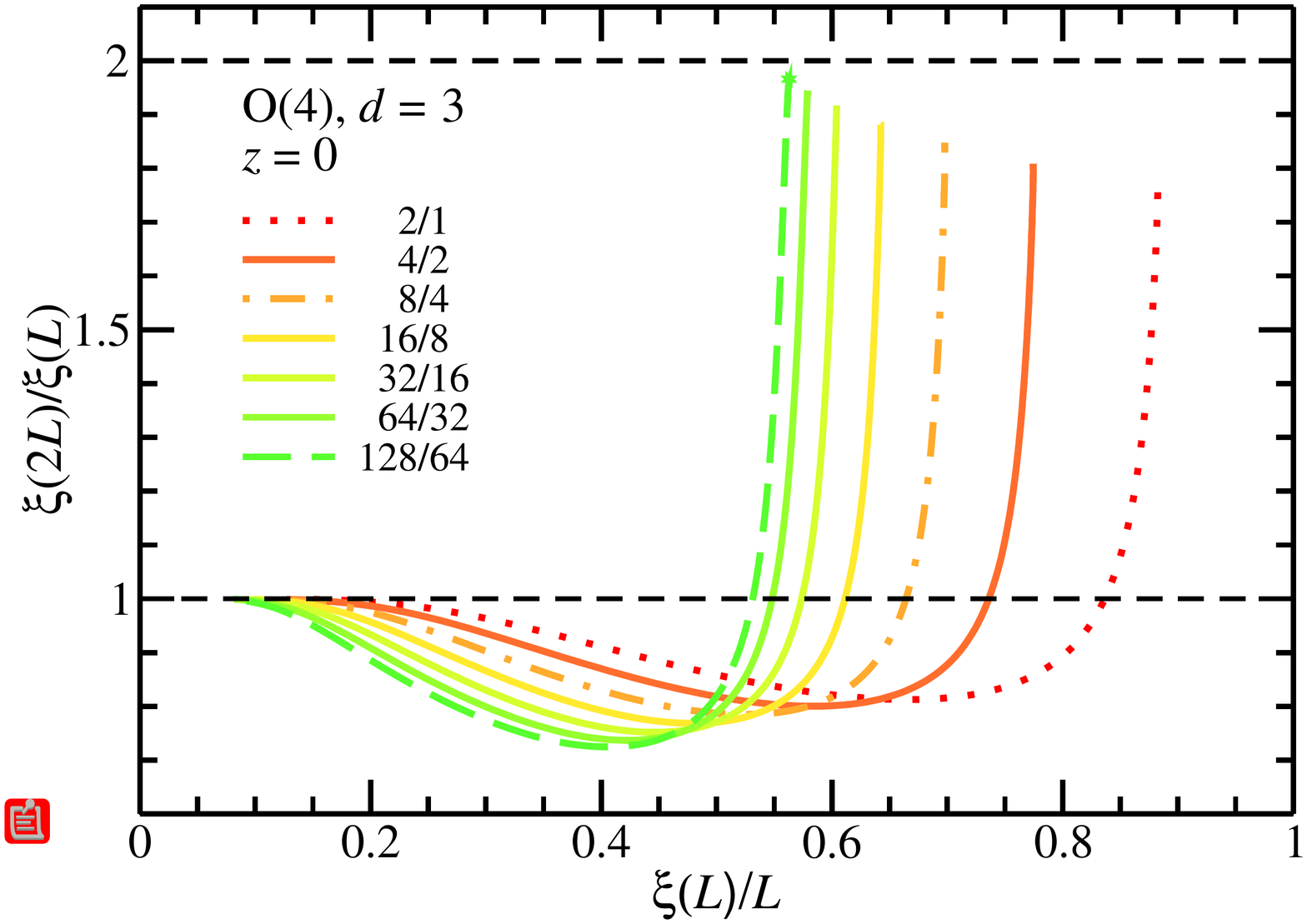}
\includegraphics[scale=0.28, clip=true]{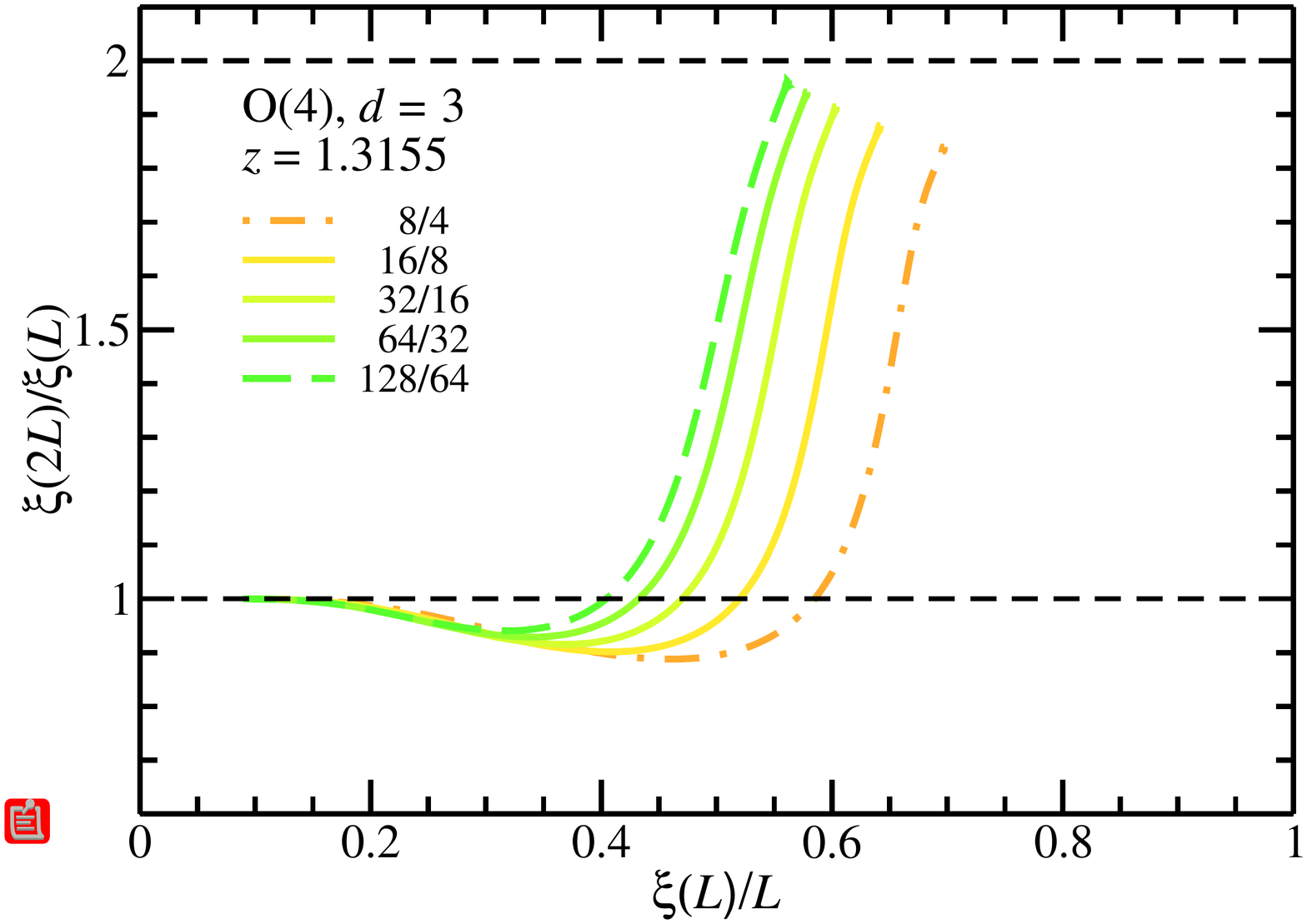}
\caption{Ratio of the finite-volume correlation lengths $\xi(2L)/\xi(L)$ for systems of the volume sizes $2L$ and $L$ als a function of the ratio $\xi(L)/L$ of the finite-volume correlation length and the system size  for the smaller system. Since the finite-volume correlation length cannot exceed the system size, for the given choice of $\xi(2L)/\xi(L)$ the ratio is bounded by $2$ from above. For large volume size ($\xi(L)/L \to 0$), the ratio must tend to unity. }
\label{fig:xiratios}
\end{figure}

Alternatively, one can also calculate the ratio of an observable for two different volume sizes $L$ and $sL$, which differ by a fixed factor $s$, and plot the results as a function of the dimensionless correlation length $\xi(L)/L$ for the smaller volume. The results of this procedure for the order parameter $M(2L)/M(L)$ and for the correlation length itself $\xi(2L)/\xi(L)$
are shown in Figs.~\ref{fig:Mratios} and \ref{fig:xiratios}. We have again chosen the values $z=0$ and $z=z_p$.
The plots are generated by keeping the scaling variable $z$ fixed while varying $h$, which changes the correlation length.
%
%
As expected, we find that for both observables the ratio ${\mathcal O}(2L)/{\mathcal O}(L)$ tends to one in the infinite-volume limit $\xi(L)/L \to 0$, where the correlation length 
for a given external symmetry breaking field becomes much smaller than the volume size and the finite-size effects disappear. 
In small volumes, corrections to the scaling behavior are unfortunately large, and a systematic extrapolation is more difficult here, since the scaling functions are given in parametric form. 
But it is clear that the results converge to a universal result for these scaling functions for large volume size, as well. 

In particular for the correlation length, there is an additional check that helps us to evaluate the convergence behavior: As we have already noticed in the discussion of Fig.~\ref{fig:QofxioverL}, the correlation length is bounded by the volume size (at apparently $\approx 0.5 L$). For growing correlation length, the ratio $\xi(2L)/\xi(L)$ in volume sizes $2L$ and $L$ must eventually  converge to $\xi(2L)/\xi(L)=2$, and with increasing volume size, we observe exactly this convergence to the expected value.

We also observe that the shapes of the scaling functions for the ratios vary with the value of the scaling variable $z$, which could also conceivably help to distinguish between different universality classes.

\clearpage

\section{Conclusions}
\label{sec:conclusions}
Using functional RG methods, we have calculated the finite-size scaling functions for the O(4) model in three dimensions for a wide range of temperature, symmetry breaking and volume size. The behavior of the scaling functions allows us to identify two distinct regions: for large values of the correlation length and small values of the finite-size scaling variable, finite-size effects are large, and the behavior of the observables is described by finite-size scaling. Above a certain value for the scaling variable, where the correlation length is sufficiently small, the results converge rapidly towards the infinite-volume behavior, and the scaling functions exhibit asymptotic behavior consistent with predictions from infinite-volume scaling.

We have expanded the effective action in terms of local interactions. As discussed in \cite{Braun:2007td}, this introduces a 
systematic error into the values for all critical exponents used in the evaluation of the results. We have used two different RG schemes in our calculation in order to assess truncation effects 
on our results, which allows us to estimate a theoretical error. In the infinite-volume limit, the results for both calculations coincide, which is also the case for very small volumes, where all
fluctuations are cut off.

Deviations from the scaling behavior in our results can be understood by RG arguments as scaling corrections, and we have checked that the behavior is consistent with that predicted by the RG. 
We have removed these corrections, which appear primarily for very small volumes and very large values of the external symmetry-breaking fields, and extrapolated from our results to the universal finite-size scaling functions. We provide parameterizations for our scaling functions separately for the finite-size scaling region at small values of the scaling variables and for the 
asymptotic region at large values of the scaling variables.

For selected values of the infinite-volume scaling variable $z$, we have confirmed that the expected scaling relations between the scaling functions for the order parameter and the susceptibility
hold for our results. We have also briefly presented some results for scaling functions in terms of dimensionless ratios of observables, as a function of the dimensionless correlation 
length $\xi(L)/L$, which allow direct comparisons of different systems in the same universality class without any need for additional normalization. 

Our results are useful for a comparison to Lattice QCD simulations. In order to perform such a comparison for the scaling of the order parameter (or the susceptibility) to 
results for the chiral condensate (or the chiral susceptibility) from Lattice QCD as a function of the volume size, a determination of the proper normalization constants for
the reduced temperature, the external symmetry breaking field and the correlation length for the QCD results is necessary. 
Once these constants are known, a direct comparison is possible. This will help to 
shed more light on the question whether the QCD phase transition falls into the $O(4)$ universality class or not. Such a detailed comparison is work in progress~\cite{BraunKlein:2008}. 
In this respect, we would like to remark that
our results at the present stage appear to suggest that -- at least with regard to a finite-size scaling analysis -- current Lattice QCD simulations might be in the asymptotic region, where the correlation length is much smaller than the lattice size and where large scaling effects cannot be observed. This is consistent with the observations previously made in \cite{Engels:2001bq}.
For a meaningful scaling analysis of Lattice QCD results in terms of $d=3$ universality classes,  the correlation length $\xi$ on the lattice has to be much larger than the length scale set by the inverse temperature.  At the same
time, it must be of the order of the spatial extension $L$ for a finite-size scaling analysis.
If these conditions cannot be satisfied simultaneously, then the universality class of the QCD phase transition cannot be determined from the comparison of Lattice QCD results to scaling behavior as it is observed in finite-volume $d=3$ models.

\begin{acknowledgments}
The authors would like to thank Daniel Litim, Tereza Mendes, and Andreas J\"uttner for useful discussions and Andreas J\"uttner for providing values for the critical exponent $\omega$ for different truncations. 
BK would like to thank TRIUMF for the hospitality during the completion of this work.
This work was supported by the Research Cluster "Structure and Origin of the Universe" and by the Natural Sciences and 
Engineering Research Council of Canada (NSERC). TRIUMF receives federal funding via a contribution agreement 
through the National Research Council of Canada.
\end{acknowledgments}

\newpage
\appendix
\section{Tables}

\begin{table}[h]
\caption{Values of the non-universal normalization constants for our parameter set used in the evaluation. Following the usual convention, they are defined from the normalization conditions $M(t, h =0) =(- t)^\beta = \left(\frac{T_c-T}{T_0}\right)^\beta$ for $h=0$ and $t<0$, and $M(t=0, h) = h^{1/\delta}=\left(\frac{H}{H_0}\right)^{1/\delta}$ for $t=0$. For details, see \cite{Braun:2007td}. Additionally, we use here $\xi(t, h =0) = \frac{1}{C_0} t^{-\nu}$. the overall scale is set by our choice $\Lambda= 10^3$ MeV.}
\label{tab:normalization}
\begin{center}
\begin{tabular}{c c c c}
\hline
\hline
$T_c$ &  $T_0$ & $H_0$ & $C_0$\\
\hline
$13.682\,368\,165\,072\,75\,$ MeV & $0.014916(5)$ MeV & $6.032(10)$ MeV$^{5/2}$& $2.393(2)$ MeV\\
\hline\hline
\end{tabular}
\end{center}
\end{table}
%
\begin{table}[h]
\caption{Values of the critical exponents used in the evaluation. We use the values determined in~\cite{Braun:2007td} in local potential approximation. See 
Ref.~\cite{Braun:2007td} for a discussion of the accuracy and effects of the approximation.}
\label{tab:critex}
\begin{center}
\begin{tabular}{c c c c c}
\hline\hline
$\beta$  & $\nu$ & $\delta$  & $\gamma$ & $\eta$\\
\hline
$0.4030(30)$ & $0.8053(50)$ & $4.9727(5)$ & $1.606(10)$ & $0$ \\ 
\hline\hline
\end{tabular}
\end{center}
\end{table}
%
\begin{table}[h]
\caption{Parameterization of the order parameter finite-size scaling function $Q_M(z, hL^{\beta \delta/\nu})$ for small values of the scaling variable $hL^{\beta\delta/\nu}$, determined from the results from $L=1-10$ fm and from $L=10-100$ fm. The last column indicates the value up to which this parameterization is approximately valid.}
\begin{center}
\begin{tabular}{ c  c  c  c  c  c  }
\hline
\hline
$z$ & $c(z)$ &  $\tau(z)$ & $(h L^{\beta \delta/\nu})$  \\
\hline
$-0.5$ & $1.708(39)$ & $0.9578(144)$ & $0.58$\\
$-0.2$ & $1.585(32)$ & $0.9536(115)$ & $0.57$\\
$-0.1$ & $1.524(25)$ & $0.9455(116)$  & $0.57$\\
$\phantom{-}0.0$ & $1.464(31)$ & $0.9371(118)$ & $0.58$\\
$\phantom{-}0.1$ & $1.405(30)$ & $0.9284(121)$ & $0.58$\\
$\phantom{-}0.2$ & $1.347(30)$ & $0.9194(125)$  & $0.58$\\
$\phantom{-}0.5$ & $1.140(28)$ & $0.8734(152)$ & $0.61$\\
\hline
\hline
\end{tabular}
\end{center}
\label{tab:QMparasmallhsmallz}
\end{table}%
%
\begin{table}[h]
\caption{Parameterization of the order parameter finite-size scaling function $Q_M(z, hL^{\beta \delta/\nu})$ for small values of the scaling variable $h L^{\beta \delta/\nu}$, determined from the results from $L=10-100$ fm. The last column indicates the approximate value of the scaling variable $h L^{\beta \delta/\nu}$ below which the parameterization is valid.}
\begin{center}
\begin{tabular}{l  l  c  c  }
\hline
\hline
\multicolumn{1}{c}{$z$} & \multicolumn{1}{c}{$c(z)$} &  $\tau(z)$ &  $(h L^{\beta \delta/\nu})$  \\
\hline
$-5.0$ & $4.472(187)$ & $1.1067(227)$ & $0.42$\\ 
$-4.0$ & $3.953(140)$ & $1.0991(191)$ & $0.43$\\ 
$-3.0$ & $3.3848(983)$ & $1.0848(158)$ & $0.45$\\
$-2.0$ & $2.6825(726)$ & $1.0444(156)$ & $0.50$\\
$-1.0$ & $2.0647(440)$ & $1.0043(122)$ & $0.54$\\
$\phantom{-}0.0$ & $1.4396(308)$ & $0.9340(121)$ & $0.59$\\
$\phantom{-}1.0$ & $0.9012(289)$ & $0.8302(179)$ & $0.56$\\
$\phantom{-}1.3155$ & $0.7890(315)$ & $0.8069(207)$ & $0.50$\\
$\phantom{-}2.0$ & $0.6453(499)$ & $0.7887(314)$ & $0.31$\\ 
\hline
\hline
\end{tabular}
\end{center}
\label{tab:QMparasmallhlargez}
\end{table}%
%
\begin{table}[h]
\caption{Parameterization $Q_M(z, hL^{\beta \delta/\nu}) = c_\infty(z) (hL^{\beta \delta/\nu})^{\tau_\infty}$ of the order parameter finite-size scaling function, for large values of the scaling variable where the behavior becomes asymptotic, determined from the results from $L=1-10$ fm and from $L=10-100$ fm. From the asymptotic properties of the scaling function, one expects that $c_\infty(z)=f(z)$ and that $\tau_\infty = 1/\delta$ is independent of $z$. For comparison, the value $f(z)$ of the infinite-volume scaling function and $1/\delta$ are also listed.
}
\begin{center}
\begin{tabular}{l  c  c  l  c   }
\hline
\hline
\multicolumn{1}{c}{$z$} & $c_\infty(z)$ & $f(z)$ & \multicolumn{1}{c}{$\tau_\infty$} & $1/\delta$ \\
\hline
$-0.5$  & $1.14303(39)$ & $1.15371$ & $0.22611(15)$  &  $\cdots$ \\ 
$-0.2$  & $1.05861(17)$ & $1.06353$ & $0.22807(7)$  &  $\cdots$  \\
$-0.1$  & $1.02603(26)$ & $1.03179$ & $0.23014(11)$  &  $\cdots$\\
$\phantom{-}0.0$ & $0.99587(25)$ & $0.99916$ & $0.23113(10)$  & $0.2011(1)$\\
$\phantom{-}0.1$ & $0.96286(33)$ & $0.96564$ & $0.23343(14)$  &  $\cdots$\\
$\phantom{-}0.2$ & $0.93380(28)$ & $0.93122$ & $0.23337(13)$  & $\cdots$ \\
$\phantom{-}0.5$ & $0.83478(19)$ & $0.82288$ & $0.23907(10)$  &  $\cdots$\\
\hline
\hline
\end{tabular}
\end{center}
\label{tab:QMparaasympsmallz}
\end{table}%
%
\begin{table}[h]
\caption{Parameterization $Q_M(z, hL^{\beta \delta/\nu})=c_\infty(z) (hL^{\beta \delta/\nu})^{\tau_\infty}$ of the order parameter finite-size scaling function  for large values of the scaling variable $h L^{\beta \delta/\nu}$, determined from the results from $L=10-100$ fm. From the asymptotic properties of the scaling function, one expects that $c_\infty(z)=f(z)$ and that $\tau_\infty = 1/\delta$ is independent from $z$. For comparison, $f(z)$ and $1/\delta$ are also listed.}
\begin{center}
\begin{tabular}{l c c c c }
\hline
\hline
\multicolumn{1}{c}{$z$} &  $c_\infty(z)$ & $f(z)$ & $\tau_\infty$ & $1/\delta$  \\
\hline
$-5.0$ &  $2.0303(8)$ & $2.0201$ & $0.2084(2)$ & $\cdots$\\ 
$-4.0$ &  $1.8745(7)$ & $1.8734$ & $0.2101(2)$ &  $\cdots$\\
$-3.0$ &  $1.7013(6)$ & $1.70843$ & $0.2116(2)$ & $\cdots$\\
$-2.0$ &  $1.5028(5)$ & $1.51793$ & $0.2131(1)$ & $\cdots$\\
$-1.0$ &  $1.2705(3)$ & $1.28915$ & $0.2145(1)$ & $0.2011(1)$\\
$\phantom{-}0.0$ &  $0.9828(1)$ & $0.99916$ & $0.2205(1)$ &  $\cdots$\\
$\phantom{-}1.0$ &  $0.6420(1)$ & $0.63228$ & $0.2403(1)$ & $\cdots$\\
$\phantom{-}1.3155$  & $0.5376(1)$ & $0.51714$ & $0.2493(1)$ & $\cdots$\\
$\phantom{-}2.0$ &  $0.3497(2)$ & $0.32481$ & $0.2667(2)$ & $\cdots$\\
\hline
\hline
\end{tabular}
\end{center}
\label{tab:QMparaasymplargez}
\end{table}%
%
\begin{table}[h]
\caption{Parameterization of the finite-size scaling function for the susceptibility $Q_\chi(z, hL^{\beta \delta/\nu})= d(z) + c_\chi(z)(hL^{\beta \delta/\nu})^{\tau(z)}$, determined from the results from $L=1-10$ fm and from $L=10-100$ fm, for small values of the scaling variable. The last column indicates the value below which the parameterization is approximately valid.}
\begin{center}
\begin{tabular}{l l l l c }
\hline
\hline
\multicolumn{1}{c}{$z$} & \multicolumn{1}{c}{$d(z)$} & \multicolumn{1}{c}{$c_\chi(z)$} & \multicolumn{1}{c}{$\tau(z)$} &  $(h L^{\beta \delta/\nu})$ \\
\hline
$-0.5$ & $0.30656(171)$ & $-1.983(888)$ & $2.557(311)$  & $0.70$\\
$-0.2$ & $0.29607(55)$ & $-0.830(67)$ & $1.893(58)$  & $0.51$\\
$-0.1$ & $0.29311(20)$ & $-0.655(16)$ & $1.702(18)$ & $0.54$\\
$\phantom{-}0.0$ & $0.29049(19)$ & $-0.535(10)$ & $1.532(14)$  & $0.56$\\
$\phantom{-}0.1$ & $0.28819(46)$ & $-0.451(16)$ & $1.380(28)$  & $0.60$\\
$\phantom{-}0.2$ & $0.28624(72)$ & $-0.391(18)$ & $1.245(36)$  & $0.63$\\
$\phantom{-}0.5$ & $0.28207(122)$ & $-0.304(10)$ & $0.952(31)$ & $0.70$\\
\hline
\hline
\end{tabular}
\end{center}
\label{tab:suscparasmallhsmallz}
\end{table}%
%
\begin{table}[h]
\caption{Parameterization of the finite-size scaling function for the susceptibility $Q_\chi(z, hL^{\beta \delta/\nu})= d(z) + c_\chi(z) (z, hL^{\beta \delta/\nu})^{\tau(z)}$, determined from the results from $L=10-100$ fm, for small values of the scaling variable.}
\begin{center}
\begin{tabular}{l l l l c}
\hline
\hline
\multicolumn{1}{c}{$z$} & \multicolumn{1}{c}{$d(z)$} & \multicolumn{1}{c}{$c_\chi(z)$} & \multicolumn{1}{c}{$\tau(z)$} &  $(h L^{\beta \delta/\nu})$ \\
\hline
$-5.0$ & $0.5382(392)$ & $-1.567(598)$ & $1.702(498)$ & $0.52$\\
$-4.0$ & $0.4887(305)$ & $-1.463(547)$ & $1.795(483)$ &  $0.52$\\
$-3.0$ & $0.4376(219)$  & $-1.325(460)$ & $1.890(444)$ &  $0.53$\\
$-2.0$ & $0.3854(138) $ & $-1.105(313)$ & $1.939(362)$ &  $0.55$\\
$-1.0$ & $0.3306(53)    $ & $-1.134(300)$ & $2.162(276)$ &  $0.53$\\
 $\phantom{-}0.0$ & $0.29028(117)$ & $-0.4455(146)$ & $1.3762(368)$ &  $0.63$\\
 $\phantom{-}1.0$ & $0.28269(128)$ & $-0.2625(27)$ & $0.6135(131)$ &  $0.68$\\
 $\phantom{-}1.3155$ & $0.2891(4)$ & $-0.2679(3)$ & $0.4946(25)$ &  $0.49$\\
 $\phantom{-}2.0$ & $0.3198(95)$ & $-0.3111(36)$ & $0.3246(161)$ &  $\approx 0.3$\\
\hline
\hline
\end{tabular}
\end{center}
\label{tab:suscparasmallhlargez}
\end{table}
\begin{table}[h]
\caption{Parameterization of the finite-size scaling function $Q_\chi(z, hL^{\beta \delta/\nu}) = c_\infty(z) (hL^{\beta \delta/\nu})^{\tau_\infty}$ for asymptotically large values of the finite-size scaling variable, determined from the results from $L=1-10$ fm and from $L=10-100$ fm. From the infinite-volume limit, we expect that $H_0\, c_\infty(z) = f_\chi(z)$ agrees with the infinite-volume scaling function for the susceptibility, and that $\tau_\infty = \frac{1}{\delta}-1$ is independent of $z$. For comparison, we list $f_\chi(z)$ and $ \frac{1}{\delta}-1$ in the table. 
}
\begin{center}
\begin{tabular}{c c c c c c}
\hline
\hline
$z$ &  $c_\infty(z)$ & $\tau_\infty$ & $1/\delta -1 $& $H_0\, c_\infty(z)$  & $f_\chi(z)$ \\
\hline
$-0.5$ &  $0.03466(14)$ & $-0.8124(22)$ & $\cdots$ & $0.2091(8)$ & $0.1604$\\
$-0.2$ &  $0.03786(13)$ & $-0.8015(18)$ & $\cdots$  & $0.2284(8)$ & $0.1831$\\
$-0.1$ &  $0.03889(12)$ & $-0.7953(15)$ & $\cdots$  & $0.2346(7)$ & $0.1919$\\
$\phantom{-}0.0$ &  $0.04230(15)$ & $-0.8168(21)$ & $-0.7989(1)$ & $0.2552(9)$ & $0.2014$\\
$\phantom{-}0.1$ &  $0.04369(14)$ & $-0.8142(19)$ & $\cdots$  & $0.2635(8)$ & $0.2116$\\
$\phantom{-}0.2$ &  $0.04498(15)$ & $-0.8089(19)$ & $\cdots$  & $0.2713(9)$ & $0.2225$\\
$\phantom{-}0.5$ &  $0.04986(13)$ & $-0.8056(15)$ & $\cdots$  &$0.3008(8)$& $0.2589$\\
\hline
\hline
\end{tabular}
\end{center}
\label{tab:suscparaasympsmallz}
\end{table}
%
\begin{table}[h]
\caption{Parameterization for the finite-size scaling function of the susceptibility $Q_\chi(z, hL^{\beta \delta/\nu}) = c_\infty(z) (hL^{\beta \delta/\nu})^
{\tau_\infty}$, determined from the results from $L=10-100$ fm, for asymptotically large values of the finite-size scaling variable. From the infinite-volume limit, we expect that $H_0\, c_\infty(z) = f_\chi(z)$, and that $\tau_\infty = \frac{1}{\delta}-1$, independent of $z$. We list $f_\chi(z)$ and $\frac{1}{\delta}-1$ for comparison.}
\begin{center}
\begin{tabular}{l l l c c c}
\hline
\hline
\multicolumn{1}{c}{$z$} & \multicolumn{1}{c}{$c_\infty(z)$} & \multicolumn{1}{c}{$\tau_\infty$} &  $1/\delta - 1$ & $H_0\, c_\infty(z)$ & $f_\chi(z)$\\
\hline
$-5.0$ & $0.01659(14)$ & $-0.9616(38)$ &  $\cdots$ & $0.1001(9)$ & $0.06010$\\
$-4.0$ & $0.01734(12)$ & $-0.9307(34)$ &  $\cdots$ & $0.1046(7)$ & $0.06855$\\
$-3.0$ & $0.01863(11)$ & $-0.8950(29)$ &  $\cdots$ & $0.1124(7)$ & $0.08060$\\
$-2.0$ & $0.02165(12)$ & $-0.8656(28)$ &  $\cdots$ & $0.1306(7)$ & $0.09942$\\
$-1.0$ & $0.02670(10)$ & $-0.8265(19)$ & $-0.7989(1)$ & $0.1611(6)$ & $0.13271$\\
 $\phantom{-}0.0$ & $0.03698(7)$ & $-0.7933(10)$ & $\cdots$ & $0.2231(4)$ & $0.20138$\\
 $\phantom{-}1.0$ & $0.05545(8)$ & $-0.8027(9)$ & $\cdots$ & $0.3345(5)$ & $0.31666$\\
 $\phantom{-}1.3155$ & $0.05746(4)$ & $-0.7872(5)$ & $\cdots$ & $0.3466(3)$ & $0.33006$\\
 $\phantom{-}2.0$ &  $0.05137(1)$ & $-0.7388(1)$ & $\cdots$ &$0.3099(1)$ & $0.32481$\\
\hline
\hline
\end{tabular}
\end{center}
\label{tab:suscparaasymplargez}
\end{table}

\clearpage

\bibliography{O-4-scaling}




\end{document}